\documentclass[aps,twocolumn,nofootinbib,preprintnumbers,10pt]{revtex4-2}
\usepackage[utf8]{inputenc}
\usepackage[svgnames]{xcolor}
\usepackage[most]{tcolorbox}
\usepackage{caption}
\tcbset{colback=PaleGreen!0!white, colframe=NavyBlue, 
	highlight math style= {enhanced, 
		colframe=red,colback=red!10!white,boxsep=0pt}
}
\usepackage{relsize}

\usepackage{amsmath}\usepackage[bottom]{footmisc}
\usepackage{braket}
\usepackage{graphicx}
\usepackage{mwe}
\usepackage[section]{placeins}
\usepackage{framed}
\usepackage{csquotes}
\usepackage{tikz}
\usetikzlibrary{decorations.markings}
\usetikzlibrary{decorations.pathmorphing}
\definecolor{shadecolor}{rgb}{0.90,0.90,0.90}
\usepackage{hyperref}
\usepackage{bbold}
\usepackage{subcaption}
\usepackage{pifont}
\usepackage{setspace}
\usepackage{amsmath, amssymb, amsthm, float, graphicx,amsfonts}
\usepackage{amsthm}
\usepackage{array, boldline, makecell, booktabs}

\theoremstyle{definition}

\hypersetup{colorlinks=true, linkcolor=DarkRed, citecolor=DarkRed,urlcolor=Indigo,linktocpage}
\def\beq{\begin{eqnarray}}\def\eeq{\end{eqnarray}}
\def\be{\begin{equation}}\def\ee{\end{equation}}
\def\bs{\begin{split}}\def\es{\end{split}}
\def\g{\gamma}

\def\s{\sigma}
\def\m{\mu}

\def\a{\alpha}

\def\b{\beta}
\def\d{\delta}

\def\G{\Gamma}
\def\l{\lambda}

\def\G{\Gamma}

\usepackage{bm}

\def\mM{{\mathcal{M}}}
\usepackage{amstext} 
\usepackage{array}   
\newcolumntype{C}{>{$}c<{$}}
\newcolumntype{L}{>{$}l<{$}} 

\begin{document}

\title{\bf  Bell inequalities in 2-2 scattering}
\author{Aninda Sinha$^{a}$\footnote{asinha@iisc.ac.in}  and Ahmadullah Zahed$^{a}$\footnote{ahmadullah@iisc.ac.in}\\
\it ${^a}$Centre for High Energy Physics,
\it Indian Institute of Science,\\ \it C.V. Raman Avenue, Bangalore 560012, India. }

\begin{abstract}{We consider Bell inequalities in 2-2 scattering of photons, gravitons, fermions and pions.  We choose measurement settings that give maximum Bell violation for maximally entangled states and calculate the relevant Bell inequalities for these processes.  For photon scattering at low energies, QED exhibits Bell violation for all scattering angles except for a small transverse region.  This leads to a fine-tuning problem. Incorporating a light axion/axion-like particle (ALP) removes the fine-tuning problem and constrains the axion-coupling--axion-mass parameters. Allowing for graviton exchange and demanding Bell violation in photon scattering, we find that the Weak Gravity Conjecture is satisfied. Quantum gravity effect on axion coupling is discussed.  For 2-2 graviton scattering, we find that CEMZ bounds allow for at most small Bell violations. Restriction on the Weinberg angle is found by demanding Bell violation in Bhabha scattering. We use recent S-matrix bootstrap data for pions and photons to study the Bell parameter in the space of allowed S-matrices. In the photon case, we study the Bell parameters as a function of energy and find support for the EFT observations. We discuss Bell parameter for pion S-matrices, which are qutrits. For pions, we find that there is a minimization of a suitable Bell parameter for S-matrices which exhibit Regge behaviour. }
\end{abstract}
\maketitle

\section{Introduction}
Einstein, Podolsky and Rosen famously argued that special relativistic locality should imply that quantum mechanics is incomplete\cite{EPR}, suggesting that it could be completed by introducing hidden variables. The ingenious Bell inequalities \cite{bell, bellreview} led to a way to experimentally test if this is possible. The violation of these inequalities signifies that there is no local hidden variable description of the underlying physics. 
These inequalities and their generalizations have led to significant theoretical and experimental progress over the last five decades \cite{bell2, nobel}. The theoretical underpinnings of the Bell inequalities for quantum field theories have been less explored \cite{bellp1, bellp2, bellp3, bellp4, home1, beatrix,Maldacena:2015bha}--see also \cite{Green:2020whw,Choudhury:2016cso, Chakraborty:2021rvy, Raju:2018zpn}. In all these contexts, in order to calculate theoretical expectations, one assumes knowledge of the Lagrangian. Our attitude in this paper will be orthogonal to such efforts. We will ask if the Bell inequalities can be gainfully employed to constrain the theory space. Specifically, we will determine the boundary between Bell violating theories and Bell non-violating theories if interesting theories lie on such boundaries. The answer will turn out to be yes!

We will study Bell inequalities in the context of 2-2 scattering, which is arguably one of the most basic processes in quantum field theory for photon($\g$), graviton($g$), fermions($f$) and pion($\pi$). We will consider EFTs, leaving the Wilson coefficients unfixed and ask what lessons we can learn about them using the Bell inequalities. There has been a substantial amount of work during the last few years which has focused on constraining Wilson coefficients using dispersion techniques--see \cite{deRham:2022hpx} for a review.

{\it Basic setup:} Our basic set up will comprise of unentangled initial states of the same species of particles whose polarizations are fixed, in four space-time dimensions. These undergo scattering and in certain situations will lead to entangled final states. We will consider the situation where we get the same species of final state particles as what was being scattered.  The entanglement will be in helicity for the massless photons and gravitons and spin for fermions. For $\gamma, g,\psi$, there will be either two helicities or two spin components and hence we can think of these states at fixed momenta to be qubits. For qubits, the Bell inequality that we will consider is the CHSH inequality \cite{CHSH} (more appropriately, an equivalent counterpart proposed by CGLMP \cite{Belld}). As we will review below, we will fix the measurement settings to be such that the Bell parameter is maximized for the maximally entangled states $\frac{1}{\sqrt{d}}\sum_{i=0}^{d-1}\vert i,i\rangle$, with $d=2$ for qubits. This ensures that the Bell parameter is single-valued in theory space as discussed in detail in the appendix. We will work with the same settings used in \cite{Belld}---in the appendix we will justify why these settings will lead to optimal bounds on the EFT parameters.

{\it Findings--EFTs:} Our question will focus on how much Bell violation (if any) do we get for all the processes described above. In other words, we will be treating the Bell parameter as a kind of ``order parameter'' in theory space. We will find several remarkable features which we summarize here:

\begin{itemize}
\item We consider $\gamma\gamma\rightarrow \gamma\gamma$ (LbyL) in an EFT where the electron has been integrated out. Ligh by light (LbyL) scattering, specifically $\gamma\gamma\rightarrow \gamma\gamma$ in a low energy effective field theory (EFT) is described by the famous Euler-Heisenberg Lagrangian \cite{Heisenberg, Karplus, dunne}, whose first few terms are given by
\be 
\begin{split}
\mathcal{L}=-\frac{1}{4}F_{\mu \nu} F^{\mu \nu}+&\frac{g_2+f_2}{16}\left(F_{\mu \nu} F^{\mu \nu}\right)^2\\
&+\frac{g_2-f_2}{16}\left(F_{\mu \nu} \tilde{F}^{\mu \nu}\right)^2+\ldots\,,
\end{split}
\ee
with $\tilde{F}_{\mu \nu}=\frac{1}{2} \epsilon^{\mu \nu \rho \sigma} F_{\rho \sigma}$ and the coefficients $f_2$ and $g_2$, called Wilson coefficients, are known up to $O(\alpha^3)$, with $\alpha=e^2/(4\pi)=1/137$ \cite{dunne}. Their ratio is given by
\be
\frac{f_2}{g_2}\bigg|_{SM}\approx -\underbrace{0.2727}_{\text{1-loop}}-\underbrace{0.0008}_{\text{2-loop}}\approx-0.274\,.
\label{eq:f2byg2bb}
\ee
For LbyL scattering, there has been a lot of recent interest in constraining such Wilson coefficients using positivity arguments \cite{Bellazzini:2019xts,vichi1, tolleyphoton, vichi2, spin1} (see e.g. \cite{bp1, bp2} for recent work exploiting non-linear unitarity to constrain the theory space.). However, none of the existing arguments can zoom into the QED value. We find that if we demand that there is Bell violation for all values of the final state angle in the centre of mass frame, then the ratio $f_2/g_2$ obtained in this manner, which we will call $(f_2/g_2)_{Bell}$ is remarkably close to the 1-loop QED (Euler-Heisenberg) value. In fact, putting in the 2-loop correction, pushes $(f_2/g_2)_{QED}$ towards $(f_2/g_2)_{Bell}$. In LbyL scattering, we find that Bell violation for all scattering angles implies 
\be\label{bvc}
0.276\lesssim{\bigg |}f_2/g_2{\bigg |}_{Bell} \lesssim 1.21\,.
\ee
The first observation is that the upper bound is quite close to the theory space bound $| f_2/g_2| \leq 1$ although the considerations used are quite different. The second and important observation is that the standard model value $0.274$ is quite close to the lower bound obtained from Bell violation. Since there are many possible MES settings, we can ask for a given $f_2/g_2$, what is the probability that we get Bell violation for a random choice of the scattering angle. When $| f_2/g_2|\gtrsim 0.276$, then the probability becomes $1$. However, for the standard model satisfying eq.\eqref{eq:f2byg2bb}, the probability dips to $0.9995$. This appears to be a  fine-tuning (or simplicity!) problem, apparently distinct in origin from the strong CP problem \cite{Peccei:1977hh}, which is solved by the axion \cite{wein,wilc, siki}.

Can this fine-tuning problem be alleviated such that the probability alluded to above changes from 0.9995 to 1?
Accounting for an axion or Axion Like Particle (ALP) allows for a region in the axion-coupling---axion-mass plane, where indeed this is possible (see fig.\eqref{fig:BellAxion}).  It could have so happened that the coupling--mass needed lay in the experimentally excluded region. Strikingly, the allowed region is going to be experimentally probed in the near future \cite{AxionSnowMass, safdi}. 

\item In the case LbyL, if we allow for a graviton exchange, and examine Bell violation, we find  
\be\label{BellWGC}
\frac{e}{m} \geq \frac{13.8\,\sigma^{-1/4}}{M_P}\,.
\ee
where $\s={\Lambda^2}/{M_P^2}\ll 1\,.
$, with $\Lambda$ being an UV cutoff and $M_p$ is the Planck mass. This is reminiscent of the Weak Gravity Conjecture (WGC) which says that when there are interactions mediated by gauge bosons (e.g. electromagnetic interactions mediated by the photon) as well as by gravitons, then gravity will be the weakest force \cite{WGC,WGCrev, tolleyphoton, vichi2, cheung}. The above condition means that $\sigma\ll 1$ will guarantee that the Weak Gravity Conjecture (WGC) is satisfied. More conservatively, we find that if the WGC does not hold then Bell violation does not occur. 

\item If we allow for both graviton and axion exchange, and examine Bell violation, we find  
\be\label{BellWGC}
g_a \gtrsim \frac{1.22 \d^{-1/2}}{M_P}\,.
\ee
where $g_a$ axion-photon coupling, $\d={\mu^2}/{m_a^2}\,.
$ with $\mu$ being energy scale of observation below the axion mass $m_a$ This bound arises so that quantum gravity effects do not overwhelm the axion contribution.

\item In a situation, where we are scattering gravitons in an EFT, we find for Einstein gravity, there is no Bell violation (since only the MHV process is allowed). If we turn on a six derivative $\alpha_3 M_P^2 R_{ab}^{\ \ cd}R_{cd}^{\ \ e f}R^{ef ab}$ term (there is no $R^2$ correction in four spacetime dimensions as it can be field redefined away), then we find  that the condition $46.71 \gtrsim \vert \alpha_3\vert M^4\gtrsim 1.37$ leads to Bell violation. However, Bell violation is never maximal. Here $M$ is the mass of the lightest higher spin massive particle. This inequality is reminiscent of the CEMZ bound \cite{CEMZ}. Recent work in \cite{chuotgrav} uses arguments based on dispersion relations to get numerical bounds. 
\end{itemize}

{\it Findings--Bootstrap:} Next we summarize the key features that we observe using the S-matrix bootstrap. Over the last few years, attempts have been made to revive the old S-matrix bootstrap program using clever numerical techniques--see \cite{Kruczenski:2022lot} for a review. This program is still in its infancy and no doubt there will be exciting progress to be made over the next several years. Nevertheless, we will take what numerical bootstrap data is readily available and use that to examine the Bell inequalities.  We begin with photons. In \cite{joaophoton}, 2-2 scattering of photons was bootstrapped by assuming that the amplitudes arise in a unitary theory where the massive particles have been integrated out. This gives a theory space which is distinguished by low energy Wilson coefficients, the dominant one being the ratio $f_2/g_2$. One of the main goals of \cite{joaophoton} was to examine bounds on the ratios of these Wilson coefficients that arise using the full non-perturbative untarity, which includes contributions from loops. It was shown that $-1\leq f_2/g_2\leq 1$ which is the same that arises from dispersive considerations \cite{Bellazzini:2019xts,vichi1, spin1}. The S-matrix data is publicly available. We use this data and focus on the S-matrices that lie on the boundary of the allowed theory space. Our main focus will be the theory space (in the notation of \cite{joaophoton}\cite{foot1}.) spanned by the normalized 12-derivative Wilson coefficient $g_4/g_2^2$ and the normalized 8-derivative Wilson coefficient $f_2/g_2$.  The QED 1-loop values give $f_2/g_2=-\frac{3}{11}=-0.\bar{27}$, while $g_4/g_2^2\sim 1/\alpha^2\gg 1$. The theory space boundary found in \cite{joaophoton} was an $O(10^{-2})$ lower bound on $g_4/g_2^2$ as a function of $f_2/g_2\in[-1,1]$. This essentially means that QED sits deep inside the allowed region and not on the boundary. However, at low energies we do not expect the higher Wilson coefficients to contribute much, and as such the EFT analysis should continue to hold even for these S-matrices, which can be thought of as strongly coupled cousins of QED. 

The main conclusion that we have for the photon bootstrap data is that we find the same EFT boundary mentioned above which separates Bell violating theories from Bell non-violating theories, valid not just at low energies but up to a scale that is roughly 30 times the electron mass. This is not unexpected for two reasons. First,  $\alpha$ runs with energies. As mentioned above, including the 2-loop value \cite{dunne} $f_2/g_2=-\frac{3}{11}-\frac{130 \alpha }{363 \pi }\approx-0.2736,$ using $\alpha=1/137$. Since $\alpha$ becomes larger at higher energies, we can expect that the value for $|f_2/g_2|$ which determines the Bell violating boundary will become bigger than $-0.2736$. The numerical S-matrix data appears to supports this expectation. Next, as we mentioned above QED is deep inside the allowed region. At higher energies, the contribution of the higher Wilson coefficients will become important.

For pions, we use the numerical data in \cite{ABPHAS} obtained following \cite{smat3, andrea}. In such a set-up, we assume that the pion scattering respects crossing symmetry and unitarity. We also incorporate the location of the first massive resonance $\rho$ meson. Then to parametrize the space of allowed S-matrices, we use the Adler zeros $s_0, s_2$. If in addition to these, we enlarge the set of constraints by allow $S$ and $D$ wave inequalities that is obeyed by Chiral Perturbation Theory ($\chi$PT) as well as respected by experimental data, then we get a more interesting space of allowed S-matrices, which was dubbed as the ``river'' in \cite{ABPHAS}. Regge behaviour (where the spin of the resonances, $J$ and the squared-masses obeyed a linear relationship) was observed in two small regions on the boundary of the theory space. One region included the $\chi$PT value and lived on the upper bank of the river while another region was distinct from $\chi$PT and lived on the lower boundary. For the pion bootstrap, we observe that there is a distinct drop in the Bell parameter (obtained using the CGLMP version, since pions are qutrits) near the $\chi$PT location. In \cite{ABPHAS}, it was pointed out that a drop in entanglement (specifically the entanglement power) is correlated with the emergence of Regge behaviour. An argument why this is plausible was presented in \cite{Beane:2021zvo}. Our findings give further support to this observation but in a clearer manner than what was possible in \cite{ABPHAS}. It appears to us that the Bell parameter is a better ``order parameter'' to distinguish theories. In \cite{Aoude:2020mlg}, entanglement minimization was used to derive the parameters of Einstein gravity.

\section{Bell inequalities in 2-2 scattering}
\begin{figure}[ht]
  \centering
  \includegraphics[width=0.6\linewidth]{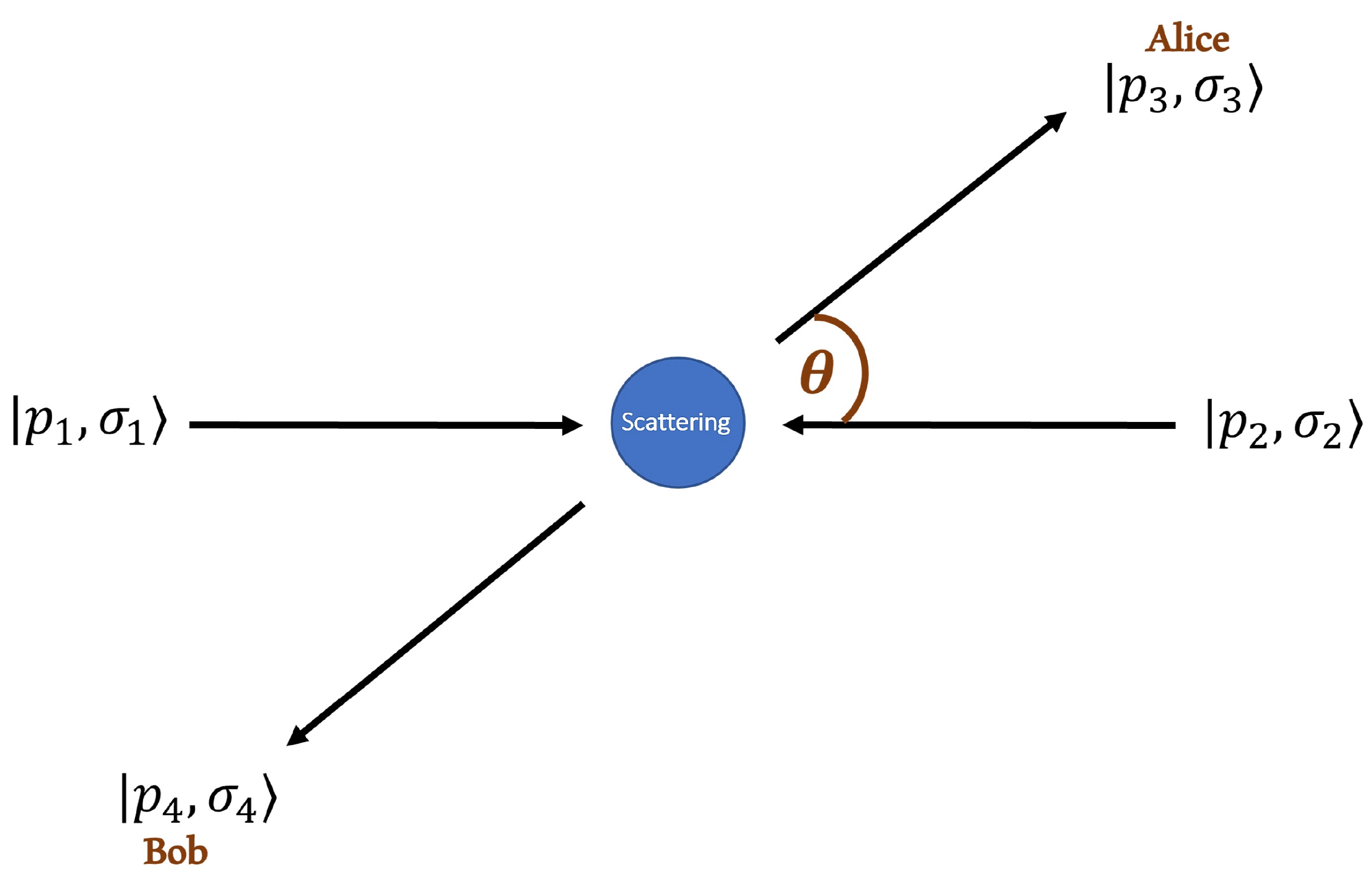}
 \caption{Two particles $| p_1, \s_1\rangle$ and $| p_2, \s_2\rangle$ scatter and one particle ($| p_3, \s_3\rangle$) reaches Alice and another ($| p_4, \s_4\rangle$) reaches Bob, where $| p_i, \s_i\rangle$ denotes a state of momentum $p_i$ and helicity $\s_i \equiv \pm$.}
\label{fig:Scattering_COM2}
\end{figure}

We briefly discuss Bell inequalities \cite{bell, CHSH, Belld} in the context of 2-2 scattering. First we will focus on qubit system. The general dimensional system will be discussed section \eqref{sec:pion}.
Our set up is illustrated in figure \eqref{fig:Scattering_COM2}. After scattering, one particle  reaches Alice (A) and another reaches Bob (B). Alice decides to measure one of two physical properties, $A_1$ or $A_2$,  taking values $0,1$. Similarly Bob can measure $B_1$ or $B_2$, taking values $0,1$. The joint probability for Alice measuring $A_1=j$ and Bob measuring $B_1=l$ is given by
$P(A_1=j, B_1=l)$. Similar interpretation follows for $P(A_1=j, B_2=m)$, $P(A_2=k, B_1=l)$ and 
$P(A_2=k, B_2=m)$. Specific combinations of the joint probabilities can be written which give an upper bound for local hidden variable theories, which can be violated in a quantum theory. We follow \cite{Belld} to define the CGLMP version of the Bell parameter:
\be
\label{I3}
\begin{split}
I &=  \Big [ P(A_1=B_1) +  P(B_1= A_2+1)  + P(A_2=B_2)\\
 +& P(B_2=A_1) \Big]-  \Big [ P(A_1=B_1-1)+  P(B_1= A_2)\\
 + & P(A_2=B_2-1) +  P(B_2=A_1-1)\Big]
\end{split}
\ee
where the probability $P\left(A_a=B_b+\right.$ $k)$ is defined using the joint probabilities via \cite{PAB},
\be\label{eq:PA=B+k}
P\left(A_a=B_b+k\right) \equiv \sum_{j=0}^{1} P\left(A_a=(j+k) \bmod 2, B_b=j\right) .
\ee
For a local deterministic hidden variable theory, one has $B_1-A_1=l-j, A_2-B_1=k-l, B_2-A_2=m-k, A_1-B_2=j-m$, which add to zero.  Thus, any three probabilities in the first line of \eqref{I3} and one in the second line, or vice versa, can hold \cite{Belld}. In such a theory, $-2\leq I\leq 2$. In a quantum mechanical theory, one finds that $| I|>2$ is possible and the maximum value of the Bell parameter $| I |_{max}=2\sqrt2$.  Bell violation in a quantum theory means
\be \label{bellv}
2< | I|\leq 2\sqrt{2}\,.
\ee
Quantum observables are hermitian operators. Since we are considering a qubit system, the general form of the operators $\hat{A}_1, \hat{A}_2$ and $\hat{B}_1, \hat{B}_2$, measured by Alice and Bob are a linear combination of Pauli operators $\sigma_1, \s_2,\s_3$. The coefficients in these combinations are called \textit{measurement parameters}. For example CGLMP \cite{Belld} choose $
\hat{A}_a=-\cos \pi \alpha_a ~\sigma_1+ \sin \pi \alpha_a ~\sigma_2\,,~
\hat{B}_b=-\cos \pi \beta_b~ \sigma_1+ \sin \pi \beta_b~\sigma_2
$
where the measurement parameters are $\alpha_1=0, \alpha_2=1 / 2, \beta_1=1 / 4, \beta_2=-1 / 4$. After we have the operators for measurements, and given the quantum states $ | \psi\rangle$ of two dimensional entangled systems, we calculate the joint probability, and hence the Bell parameter $I$. For example, the choice in \cite{Belld} gives the value of $I$ to be $2\sqrt{2}$ for the maximally entangled state $(|0,0\rangle+| 1,1\rangle)/\sqrt{2}$. There are multiple measurement parameters that can give the maximum value of the Bell parameter $| I|_{max}=2\sqrt2$.\\

\textbf{MES setting}: We define an \textit{MES setting} as a particular set of measurement parameters that give $| I| =2\sqrt{2}$ for a maximally entangled state. Why the MES settings take a distinguished position is explained in the supplementary material.

\section{The 2-2 scattering amplitudes}
In this section we introduce the basic notion for 2-2 scattering amplitudes. Let the initial state and final state be $\left|p_1, \l_1 ; p_2, \l_2\right\rangle$ and $\left|p_3, \l_3 ; p_4, \l_4\right\rangle$ respectively, where $\l_1, \l_2, \l_3$ and $\l_4$ helicity indices, take the values $\l_i=+h,-h$, with $h=1$ for photons and $h=2$ for gravitons. We will assign $+h\to 1, -h\to 0$, for notational convenience. For the pion scattering the $\l_i$ are iso-spin indices, take values $0,1,2$. The S-matrix (in-out convention) is defined as
\be
\begin{split}
&\left\langle p_3, \l_3 ; p_4, \l_4|S| p_1, \l_1 ; p_2, \l_2\right\rangle\\
&=1+i \delta^4\left(p_1+p_2-p_3-p_4\right) \mathcal{M}_{\l_1 \l_2}^{\l_3 \l_4}(s, t, u),
\end{split}
\ee
where the $s, t, u$ are usual Mandelstam variables such that $s+t+u=0$ for photons or gravitons, while $s+t+u=4m^2$ for pions and electrons. We will be interested in situations where the initial states have undergone scattering and hence, $\mathcal{M}_{\l_1 \l_2}^{\l_3 \l_4}(s, t, u)$ is the object of interest.

\subsection{Photons and gravitons}

There are total sixteen photon or graviton transition scattering amplitudes. Since we are considering scattering of identical particles and we assume parity symmetry, there are only five distinct center of mass amplitudes\cite{joaophoton}. We will denote then as 
\be
\begin{aligned}
&\Phi_1(s, t, u) \equiv \mathcal{M}_{++}^{++}(s, t, u), ~\Phi_2(s, t, u) \equiv \mathcal{M}_{++}^{--}(s, t, u),  \\
&\Phi_3(s, t, u) \equiv \mathcal{M}_{+-}^{+-}(s, t, u),~\Phi_4(s, t, u) \equiv \mathcal{M}_{+-}^{-+}(s, t, u),  \\
&\Phi_5(s, t, u) \equiv \mathcal{M}_{++}^{+-}(s, t, u) .
\end{aligned}
\ee
Due to crossing symmetry only the amplitudes are independent are $\Phi_1, \Phi_2$, and $\Phi_5$ , while the $\Phi_3(s, t, u)$ and $\Phi_4(s, t, u)$ can be related to $\Phi_1(s, t, u)$ as
$$
\Phi_3(s, t, u)=\Phi_1(u, t, s), \quad \Phi_4(s, t, u)=\Phi_1(t, s, u)\,.
$$
The rest of the helicity configurations are related to $\Phi_i$'s   as follows
\be\label{eq:ampphoton}
\left(
\begin{array}{cccc}
 \mM_{+-}^{+-} & \mM_{+-}^{++} & \mM_{--}^{+-} & \mM_{--}^{++} \\
 \mM_{+-}^{--} & M_{+-}^{-+} & M_{--}^{--} & M_{--}^{-+} \\
 M_{++}^{+-} & \mM_{++}^{++} & \mM_{-+}^{+-} & \mM_{-+}^{++} \\
 \mM_{++}^{--} & \mM_{++}^{-+} & \mM_{-+}^{--} & \mM_{-+}^{-+} \\
\end{array}
\right)=\left(
\begin{array}{cccc}
 \Phi _3 & \Phi _5 & \Phi _5 & \Phi _2 \\
 \Phi _5 & \Phi _4 & \Phi _1 & \Phi _5 \\
 \Phi _5 & \Phi _1 & \Phi _4 & \Phi _5 \\
 \Phi _2 & \Phi _5 & \Phi _5 & \Phi _3 \\
\end{array}
\right)\,.
\ee

\subsection{Fermions}

In this section we will consider 2-2 fermion scattering, namely for illustration we consider Bhabha scattering of electron-positron pair with photon, Z-boson and graviton exchange. The amplitude for a generic exchange with the explicit spin index, can be written as
\begin{small}
\be\label{eq:bhaba}
\begin{split}
\mM_{s_1,s_2}^{s_3,s_4}&= \bar{v}^{s_2}(p_2)~\G^{\{A\}}~ u^{s_1}(p_1)~ E_{\{A\},\{B\}} ~\bar{u}^{s_3}(p_3)~\G^{\{B\}}~v^{s_4}(p_4)\\
&-(2\leftrightarrow 3)
\end{split}
\ee
\end{small}
where $\{A\}, \{B\}$ are proxy for summed dummy index (there can be more than one).

For photon exchange the coupling and propagator is given by is given by 
\be\nonumber 
\G^{\{A\}}=i  e \gamma^\mu,~~ \G^{\{B\}}=i  e \gamma^\nu,~~ E_{\{A\},\{B\}}=\frac{\eta_{\mu ,\nu}}{s}\,.
\ee

For Z-boson exchange the coupling and propagator is given by is given by 
\be\nonumber
\begin{split}
&E_{\{A\},\{B\}}=\frac{\eta_{\mu ,\nu}}{s-M_z^2}\,,~\G^{\{A\}}=i \frac{g}{\cos \theta_W} \gamma^\mu\left(g_V^f-g_A^f \gamma^5\right),\\
& \G^{\{B\}}=i \frac{g}{\cos \theta_W} \gamma^\nu\left(g_V^f-g_A^f \gamma^5\right),
\end{split}
\ee
where $g=\frac{e}{\sin{\theta_w}},~ g_V^f=-1/4,~ g_A^f=-1/4+\sin^2\theta_w$, with $\theta_w$ being the Weinberg angle.

For graviton exchange the coupling is given by is given by \cite{Santos}
\be\nonumber
\begin{split}
&\G^{\{A\}}=-\frac{\mathrm{i} \kappa}{4}\left(\gamma^\mu p_1^\nu+p_2^\mu \gamma^\nu\right),\,\G^{\{B\}}= -\frac{\mathrm{i} \kappa}{4}\left(\gamma^\alpha p_3^\rho+p_4^\alpha \gamma^\rho\right) ,\\
&E_{\{A\},\{B\}}=  \frac{1}{2s}\left(g_{\mu \alpha} g_{\nu \rho}+g_{\mu \rho} g_{\nu \alpha}-g_{\mu \nu} g_{\alpha \rho}\right)\,,
\end{split}
\ee
with $\kappa=\sqrt{8\pi G}$.
We tabulate the amplitudes for each little group index in appendix in various limits of energy and angles, which will be frequently used in the upcoming sections.

\section{Bell inequality in 2-2 photon scattering}
Photons have two helicities $+$ and $-$. We will consider entanglement among the helicity states for 2-2 photons scattering. In our considerations, we fix the incoming two photons helicity to be $+ +$ (or $--$). After the scattering takes place, the final helicities can be $++,+-, -+, --$ and we will denote the transition amplitudes $\mathcal{M}_{++}^{++},~ \mathcal{M}_{++}^{+-}, ~\mathcal{M}_{++}^{-+},\text{ and } \mathcal{M}_{++}^{--}$ respectively, are functions of $s, t, u$ Mandelstam variables such that $s+t+u=0$. The subscript refers to the in-state helicities while superscript refers to the out-state helicities. We refer the reader to the supplementary material for a discussion on S-matrices of 2-2 photon scattering. At low energies, the contributions $\mathcal{M}_{++}^{-+}\approx 0 , ~ \mathcal{M}_{++}^{+-}\approx 0$ are negligible. Let us denote $\mathcal{M}_{++}^{++}=\Phi_1, \mathcal{M}_{++}^{--}=\Phi_2$. The final state works out to be
$
| \psi\rangle={\mathcal N}(\Phi_1|0,0\rangle+\Phi_2| 1,1\rangle),
$ with ${\mathcal N}=(| \Phi_1| ^2+| \Phi_2| ^2)^{-1/2}$.
We shall calculate the Bell parameter $I$ for different measurement parameters. As explained above, our choice is the MES parameters. One particular case of MES parameters, given in \cite{Belld} and reviewed above, leads to
\be
I=I^{++}=I^{--}=\frac{2 \sqrt{2} \left(\Phi _1 \Phi _2^*+\Phi _2 \Phi _1^* \right)}{\left |  \Phi _1\right | ^2+\left |  \Phi _2\right | ^2}\,.
\ee
The superscripts on $I$ indicate polarizations of the incoming photons.
For a general measurement setting, the expression for $I$ can be found in the supplementary material.
In an EFT,  the low energy transition amplitudes $\Phi_1, \Phi_2$ are real functions in general and can be written in terms of the Wilson coefficients \cite{joaophoton} (ignoring graviton exchange for now) 
$
\Phi_1(s, t, u) =g_2 s^2+\mathcal{O}\left(s^3\right) ,
\Phi_2(s, t, u) =f_2\left(s^2+t^2+u^2\right)+\mathcal{O}\left(s t u\right)\,
$. The condition \eqref{bellv} leads to
\be\label{r1byr2}
\sqrt{2}-1 \leq \left| \frac{\Phi_1}{\Phi_2}\right|\leq \sqrt{2}+1 \,.
\ee
Even though equation \eqref{r1byr2} is derived using the measurement parameters in \cite{Belld}, a crucial point is that this is the optimal bound on the ratio $| \frac{\Phi_1}{\Phi_2}|$ for any MES setting. The derivation is given in the supplementary material. 

At low energies we have
$
\left| \frac{\Phi_1}{\Phi_2}\right|=2 | f_2/g_2 |  (1-\sin ^2\frac{\theta }{2}+\sin ^4 \frac{\theta }{2})\,,
$
for a scattering angle $0\leq \theta\leq \pi$. The $\theta$ dependent part has maximum value $1$ and minimum value $\frac{3}{4}$. This translates to
\be\label{bvc2}
0.276\approx \frac{2}{3}(\sqrt{2}-1)\leq \left | \frac{f_2}{g_2}\right | \leq \frac{\sqrt{2}+1}{2}\approx 1.21 \,.
\ee

\subsection{Including the graviton}
Let us consider graviton exchange in LbyL scattering. The transition amplitudes now are given by $\Phi_1\simeq s^3/(M_P^2 t u)+\Phi_{1,SM}$, $ \Phi_2 \simeq \Phi_{2,SM}$. In this case $\mathcal{M}_{++}^{+-}(s, t, u)=\mathcal{M}_{++}^{-+}(s, t, u)=\Phi_5 \simeq \alpha\left(s^2+t^2+u^2\right)/{(360 \pi m_e^2 M_P^2)}$, where the subscript $SM$ indicates what we have been using so far and $M_P$ is the Planck mass. 
The Bell parameter $I$ for the measurement settings in \cite{Belld} and used so far works out to be
$
I=I^{++}=I^{--}=2 \sqrt{2} \left(\Phi _1 \Phi _2^*+\Phi _2 \Phi _1^* \right)/(\left \vert  \Phi _1\right \vert ^2+\left \vert  \Phi _2\right \vert ^2+2\left \vert  \Phi _5\right \vert ^2)\,.
$
Now $I$ is a function of $\tilde{e}\equiv\frac{e}{m} M_P$, the scattering angle $\theta$ and $s$. Due to the graviton pole at $t=0$, Bell inequalities are obeyed in the forward limit.  Interestingly, now there is a critical $\tilde{e}_c$ below which there is no violation for $0\leq \theta\leq \pi$. 
An approximate formula for critical value $\tilde{e}_c$ at $s=\Lambda^2$ is 
$
\tilde{e}_c\approx 13.8\,\sigma^{-1/4},\text{ where }\s={\Lambda^2}/{M_P^2}\ll 1\,.
$
We see that lower the energy,  bigger the critical value $\tilde{e}_c$. To have Bell violation for some scattering angle $\theta$, we must have 
\be\label{BellWGC}
\frac{e}{m} \geq \frac{\tilde{e}_c}{M_P}\,.
\ee
\eqref{BellWGC} is reminiscent of the weak gravity conjecture \cite{WGC, WGCrev} which states that $\frac{e}{m} \geq \frac{\sqrt{2}}{M_P}
$. If the weak gravity conjecture is not satisfied then the Bell inequality is satisfied.  

\subsection{The need for an axion/ALP}
As stated above, the QED 1-loop answer for $f_2/g_2=-\frac{3}{11}\approx-0.2727$ and after adding the 2-loop results \cite{dunne}, one has
$
f_2/g_2\approx-\frac{3}{11}-\frac{130 \alpha }{363 \pi }\approx-0.2736\,, 
$
which is closer to saturating the Bell violation condition in eq.(\ref{bvc})! One might wonder if the three-loop contribution will push it further. Unfortunately, this has not been calculated so far \cite{dunne} but it is possible to argue that this will not affect $f_2/g_2$ at the third decimal place, which is where the difference with eq.(\ref{bvc2}) lies. For three loop QED, we expect to have $-\frac{3}{11}-\frac{130 \alpha }{363 \pi }-c \alpha^2$, where $c$ is at most an $O(1)$ number, in keeping with the two-loop result, and will contribute at most to the 4th decimal place.

\begin{figure}
 \centering
  \includegraphics[width=0.7\linewidth]{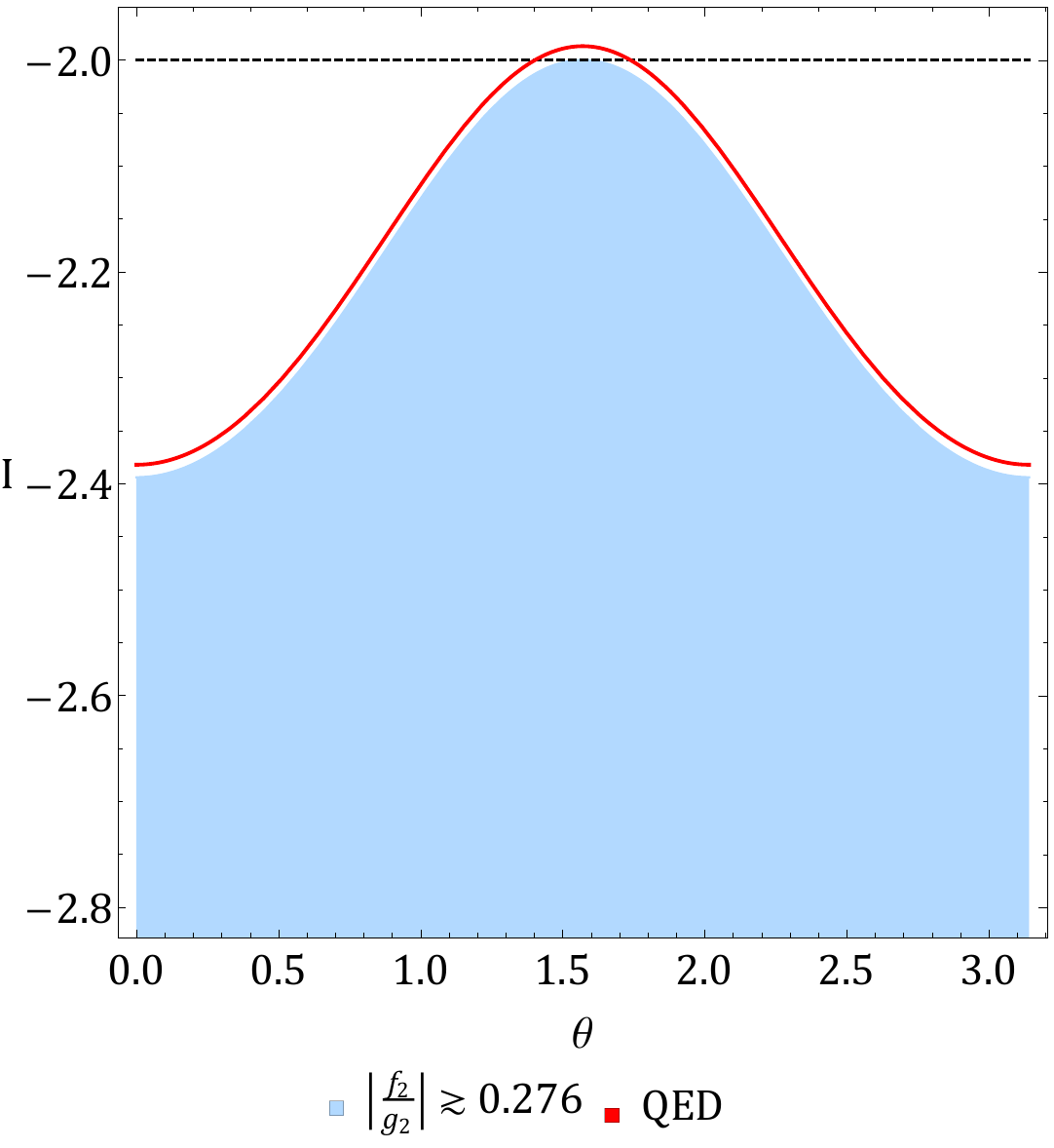}
 \caption{Bell parameter $I$  vs  the scattering angle $\theta$ for the settings in \cite{Belld}. Standard model (red) violates the Bell inequality for all $0\leq \theta\leq \pi$ except for a tiny region around $\theta=\frac{\pi}{2}$.}
\label{fig:chi_photon}
\end{figure}

\begin{figure}
  \centering
  \includegraphics[width=0.7\linewidth]{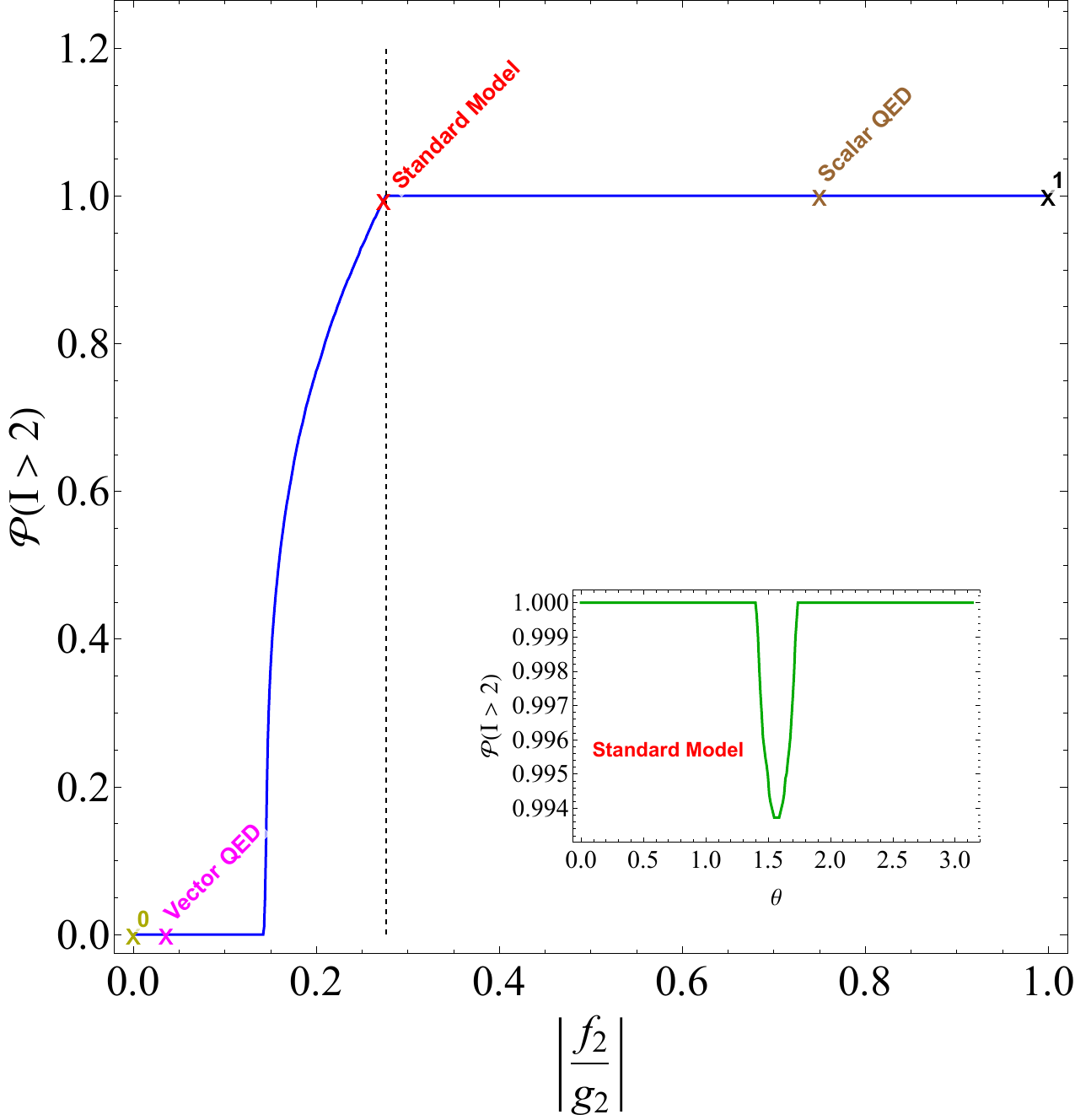}
\caption{\textbf{Main:} The probability of Bell violation in transverse direction ($\theta=\pi/2$) for random MES parameters as a function of $| f_2/g_2|$. $| f_2/g_2|\approx 0.276$ is indicated by a black dashed line while the standard model is indicated in red. \textbf{Inset:} The probability of Bell violation in the standard model  for random MES parameters \cite{supple} as a function of $\theta$. }
\label{fig:simulation}
\end{figure}
Any additional contribution from the standard model will not change the result. To see this, simply note that on dimensional grounds, we expect these contributions to be suppressed by $(m_e/M)^4$ where $M$ is the mass of the additional particle incorporated in the discussion. Such a particle can be a lepton, W-boson \cite{wcom} or pion  \cite{Engel:2012xb}. The muon and pion both will be suppressed by $10^{-10}$ while the other contributions will be more so. Hence none of SM particles can change the $f_2/g_2$ at the  $3$rd decimal place. 

In fig.(\ref{fig:chi_photon}), we have shown a zoomed plot for $I$ versus the scattering angle for the settings in \cite{Belld}. As is clear, except for a small region near $\theta=\pi/2$, $| I|\geq 2$ in the standard model. We can reinterpret the situation differently. Let us imagine experimentalists making Bell correlation measurements with any of the MES settings and at some randomly chosen $\theta$. What is the probability of obtaining Bell violation? Fig. (\ref{fig:simulation}) addresses this question. The region of interest is the transverse region as away from this region, Bell violation happens for any $\theta$. Various theories are indicated in the figure. Born-Infeld action in string theory and supersymmetric QED have $f_2/g_2=0$. Except for the standard model, all theories shown either have probability zero or probability unity. The inset figure makes it clear that for the standard model, where the main contribution comes from QED, Bell violation happens for all $\theta$ except near $\theta=\pi/2$. The probability of finding Bell violation for some randomly chosen $\theta$ in the standard model is simply the area under the green curve divided by $\pi$. This works out to be  approximately $0.9995$. We can also consider an entangled initial state $\cos p \vert 1,1\rangle+\sin p \vert 0,0\rangle$ with $0\leq p\leq \pi/2$. Here the probability of finding Bell violation for some initial state of the above form and at a randomly chosen scattering angle turns out to be  $0.999998$.

\begin{figure}[ht]
  \centering
  \includegraphics[width=0.8\linewidth]{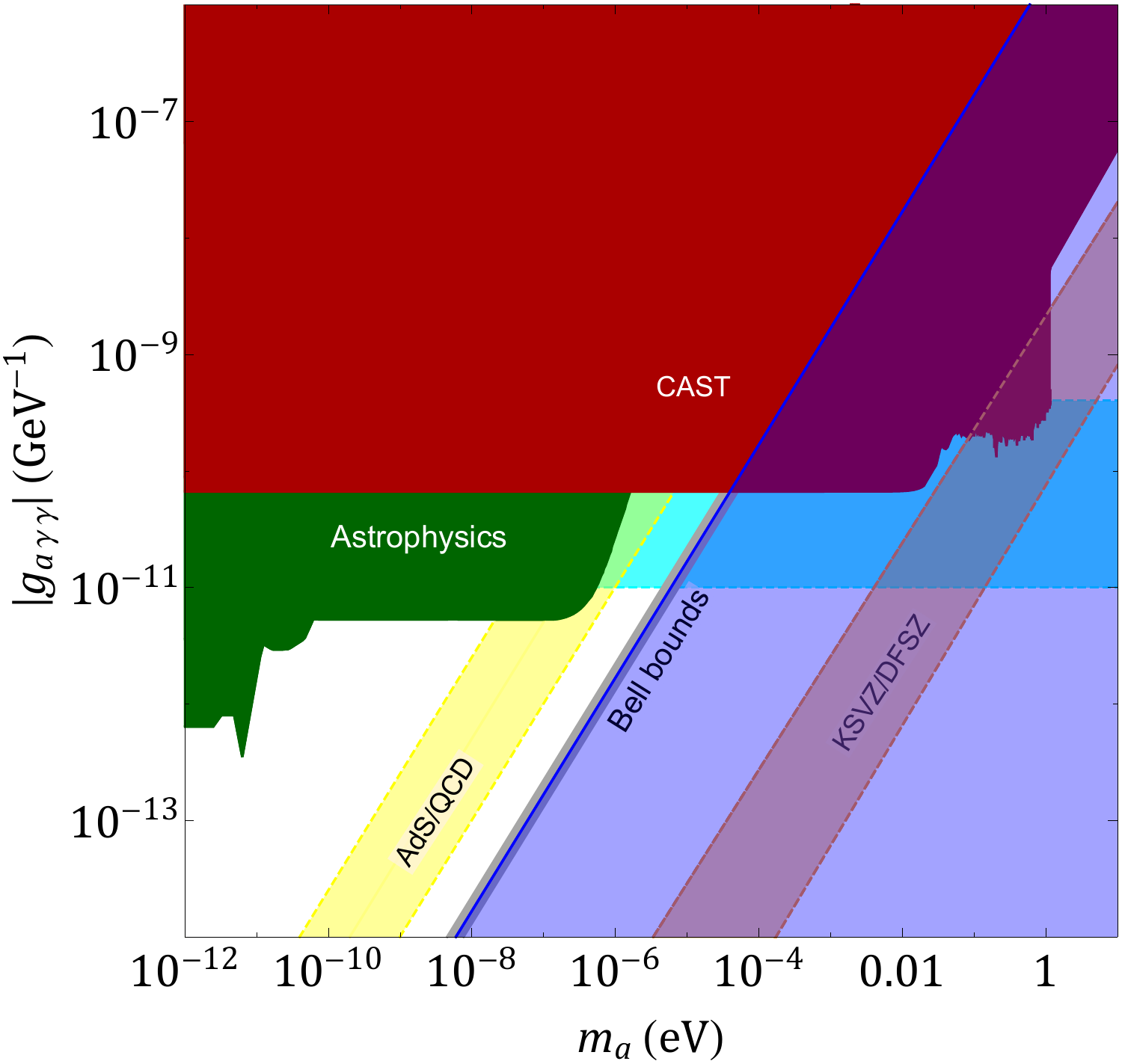}
 \caption{$\vert g_{a \g\g}\vert  ~(\text{GeV}^{-1})$ vs $m_a ~(\text{eV})$ constraints. The  CAST experiment exclusion region is given in red; the green exclusion region comes from astrophysical observations of gamma rays from SN87A, NGC1275 \cite{AxionSnowMass}. The orange band is the KSVZ/DFSZ QCD axion which solves the strong CP problem. The blue horizontal band indicates possible allowed parameter space which can explain HB star luminosity and high energy gamma transparency of the universe. The yellow band is a QCD axion obtained by integrating out a magnetically charged heavy fermion \cite{ringwald}.  Bell saturation gives the narrow gray band; an axion in the white region will remove the fine-tuning described in the text.}
\label{fig:BellAxion}
\end{figure}

Can this probability be made unity to avoid this apparent fine-tuning? From  table-1 in supplementary material, this appears to need either the pseudoscalar, or parity-odd spin-2. The pseudoscalar is the simplest and most well-motivated possibility as it corresponds to the axion/ALP. In the low energy EFT, the contribution of a single ALP to the Euler-Heisenberg Lagrangian is \cite{AxionFig3}
$
\mathcal{L}_{a, EFT}=\frac{g_{a \g \g}^2}{32 m_a^2} \left(F_{\mu \nu} \tilde{F}^{\mu \nu}\right)^2
$
where $g_{a \g \g}$ is the ALP-photon coupling and $m_a$ is the mass of the ALP. The contribution to ${g_2},~{f_2}$ is given by
\be
\frac{f_2}{g_2}\approx -\frac{3}{11}-\frac{130\a}{363\pi}-\frac{90 g_{a\g\g}^2 m_e^4}{121 \a^2 m_a^2}\,.
\ee
Now in order to saturate the $\frac{f_2}{g_2}\approx-0.276$, we must have \cite{spin2oddv2} 
\be\label{BellAxion}
\left| \frac{g_{a \g\g} ~(GeV^{-1})}{m_a ~(eV)}\right | \sim 1.66\times 10^{-6}\,.
\ee
The situation is explained figure \eqref{fig:BellAxion} along with the experimental data available \cite{AxionSnowMass}.


Remarkably, \eqref{BellAxion} allows for a parameter region which is going to be probed in near-future experiments \cite{AxionSnowMass}! If the axion saturates $f_2/g_2\approx -0.276$, then it should lie on the gray band indicated in  fig.\eqref{fig:BellAxion}. An axion in the white region can explain Bell violation for all $\theta$. Strictly speaking, so that the axion contribution does not overwhelm the one-loop answer, a narrower band close to the gray band should be considered. Such an axion can potentially explain the horizontal band (HB) star luminosity  \cite{hb} and high energy gamma transparency of the universe \cite{gammauni}. 

Can the axion/ALP that we are predicting be the one that solves the strong CP problem? In the fig.\eqref{fig:BellAxion}, the standard QCD axion is indicated by the orange band, which is in the region incapable of explaining any putative Bell violation at $\theta=\pi/2$. However, for phenomenological considerations \cite{ringwald} considered a model with an enhanced axion-coupling that arises due to integrating out a magnetically charged heavy fermion. Essentially, the enhancement is due to the Dirac charge quantization condition whereby the magnetic charge is inversely proportional to $\alpha$. This leads to the yellow band, which was dubbed as AdS/QCD in \cite{ringwald}, as the strong coupling in the calculation was restricted using AdS/QCD. Our prediction is close to this band. Such an axion is indeed capable of solving the strong CP problem. 

\textbf{Quantum gravity effects:}
The transition amplitudes now are given by $\Phi_1\simeq \Phi_{1,(\g,a)}+\frac{ s}{M_P^2 \sin^2(\theta)}$, $ \Phi_2 \simeq \Phi_{2,(\g,a)}$, where $M_P$ is the Planck mass and the subscript $(\g,a)$ indicates contribution from Photon and Axion. Note that $\Phi_{1,(\g,a)}\approx \# s^2$ in the low energy limit. Hence for a fixed $\theta$ the gravity term dominates. For simplicity we will fix $\theta=\pi/2$. 

The Bell parameter $I$ for the measurement settings in \cite{Belld} and used so far works out to be
$$
I=I^{++}=I^{--}=1.98899-\frac{1.41413 }{g_2^{(QED)} M^2_P s}+\frac{3.8388 g_a^2 m^4_e}{\a^2 m^2_a}\,.
$$
We expect that the axion contribution will dominate over quantum gravity effects, hence we find
\be 
g_a \gtrsim 1.22 \frac{\d^{-1/2}}{M_P}\,,
\ee
at a energy $s=\d\times m_a^2$, in order to have a third decimal place change in $I$. For example if $\d=0.1$, we have $g_a \gtrsim 10^{-18}\text{ GeV}^{-1}$. Similar observation was made in \cite{gaMP}.

\section{Bell inequality in 2-2 graviton scattering}
For the case of graviton, the low energy the amplitudes $\Phi_1, \Phi_2$ and $\Phi_5$ can be written as \cite{bern, chuotgrav}
\be
\begin{aligned}
\Phi_1(s, t, u) &=s^4 \left(\frac{8 \pi  G}{s t u}+\frac{2 \pi  \beta _3^2 G t u}{s}-\frac{\beta _{\phi }^2}{s}+g_2+g_3 s\right)\\
&+\mathcal{O}\left(s^6\right) \\
\Phi_2(s, t, u) &=s t u \left(40 \pi  \beta _3 G-3 \beta _{\phi }^2\right)+f_0\\
&+f_2\left(s^2+t^2+u^2\right)+\mathcal{O}\left(s^4\right) \\
\Phi_5(s, t, u) &=s^2 t^2 u^2 \left(\frac{4 \pi  \beta _3 G}{s t u}+h_3\right)+\mathcal{O}\left(s^7\right)\,.
\end{aligned}
\ee
We can consider all the higher derivative terms to be turned off such that only $\b_3$ and $G$ survives. This gives us 
\begin{small}
\be 
I^{++}=I^{--}=\frac{80 \sqrt{2} \beta _3 s^2 t^2 u^2 \left(\beta _3^2 t^2 u^2+4\right)}{\beta _3^4 s^4 t^4 u^4+8 \beta _3^2 s^4 t^2 u^2+16 s^4+408 \beta _3^2 t^4 u^4 }\,.
\ee
\end{small}
Note that Bell parameters are zero for $s=0$ as well as $\theta=0$. We find that for any finite $\chi,s$ such that $t/s=\chi=$finite, we find that the above  has maximum $I^{++}=I^{--}=\frac{20}{\sqrt{51}}=2.80056$ {(for example $\cos(\theta) =-0.44, \beta_3=2.43,s=1.89$)} and minimum is $-\frac{20}{\sqrt{51}}=-2.80056$ {(for example $\cos(\theta) =-0.28,\beta_3 =-4.67,s=1.09$)}.
\begin{figure}
  \centering
  \includegraphics[width=0.8\linewidth]{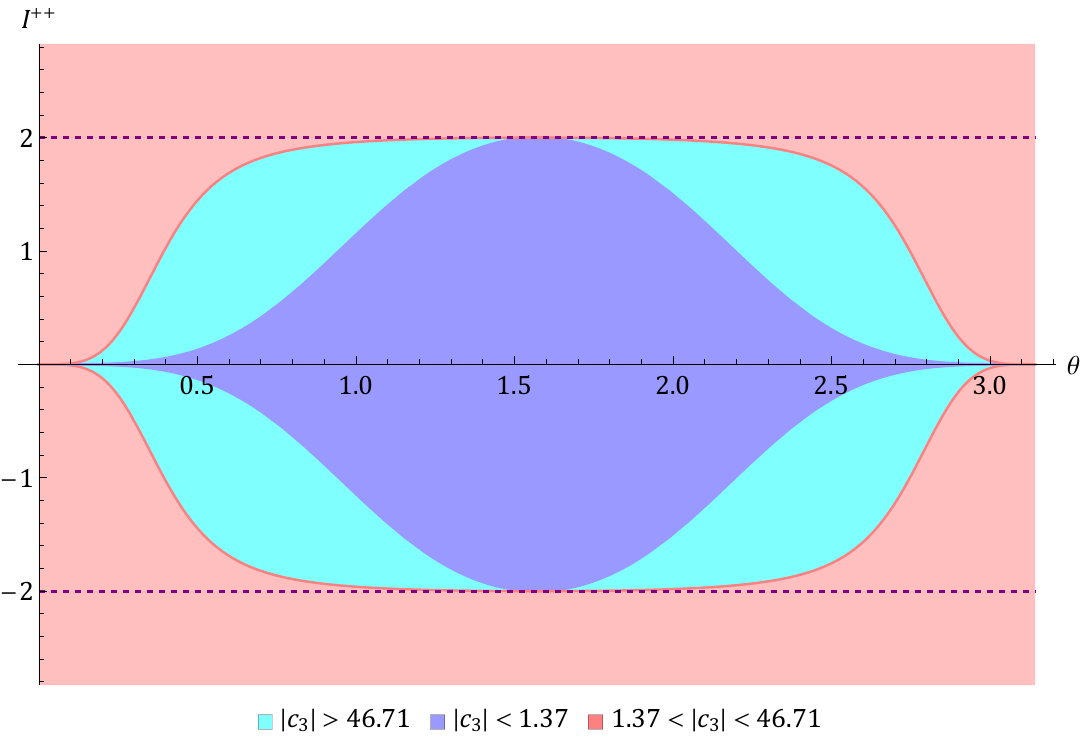}
 \caption{$I^{++}$  vs  $\theta$ for graviton in leading order in EFTs. For $1.37<|c_3|<46.71$ there will be Bell violation.}
\label{fig:chi_graviton}
\end{figure}

{We can write $\beta_3=c_3/M^4$, where $M$ is the mass of the lightest higher spin massive particle. We can compute the Bell inequalities for $s=M^2$. For physical values of the angles $0\leq \theta\leq \pi$, we find $-1.37<c_3<1.37$ leads to no Bell violations. There can be large values $|c_3|\gtrsim 46.71$ which will again give no violation. Bell violation happens only in the window $1.37\lesssim c_3\lesssim 46.71$ and $-1.37\gtrsim c_3\gtrsim-46.71$, and away from the forward/backward limit. The situation is depicted in figure \eqref{fig:chi_graviton}. If we demand Bell violation then we must have $1.37<|c_3|<46.71$, which implies that the CEMZ \cite{CEMZ} condition holds, namely $\beta_3\approx \frac{\mathcal{O}(1)}{M^4}$. The take-home statement is then ``\textit{Bell violation condition implies CEMZ like condition}" in this case.} 
Further we  find that for finite $t,s$ has maximum $I^{-+}=I^{+-}=\sqrt{2}$ and minimum is $0$. For tree level superstring theory, it turns out\cite{foot2} that $I^{++}=0=I^{-+}$.

\section{Bell inequality in Bhabha scattering}

Let us now consider Bhabha scattering $e^+e^-\rightarrow e^+ e^-$ with the initial state polarizations $++$ or $--$ allowing for photon and Z-boson (with mass $M_Z$) exchange at low energies. We find near $s\sim 4m_e^2$: $
I=I^{++}=I^{--}=-2 \sqrt{2}+\mathcal{O}\left((s-4m_e^2)^{2}\right),
$ so we always have Bell violation for all angles. For photon and graviton exchange the answer is same (for photon exchange see also \cite{Beck:2022mas}). Interestingly near $s=M_Z^2$, we find that
$
I\approx -2\sqrt{2}\sin^2 \theta/((4\sin^2 \theta_W-1)^2+1),
$
where $\sin \theta_W$ is the Weinberg angle and we have used $M_Z/m_e\gg 1$. In order to have Bell violation for some physical scattering angle $0\leq \theta\leq \pi$, one must have $0.09\lesssim\sin^2 \theta_W\lesssim 0.41$  and for maximal violation $\vert I\vert =2\sqrt{2}$ one must have $\sin^{2}\theta_W=0.25$, which is very close to the experimental value $ 0.223$. Maximal concurrence leads to the  value $\sin^{2}\theta_W=0.25$ \cite{Cervera-Lierta:2017tdt}. This agreement is expected, since for qubits there is a one-to-one correspondence between maximum Bell violation and maximum concurrence.

In the case of Bhabha scattering with graviton exchange since the exchange pole is a massless pole, we don't see any noticeable change in the Bell parameters. The reason is that to excite the pole contribution we have to probe the energy regime around the pole, which is $s=0$. This is not a physical regime as  $s\geq 4m_e^2$ in this case.
We always find $I_3^{+-}=I_3^{-+}=0$ in the forward limit and $I_3^{+-}=I_3^{-+}=O\left(s-4m_e^2\right)^2$ in the low energy limit.

\section{Bell inequality in 2-2 pion scattering}\label{sec:pion}
Since pion system is a qutrits system (3-dimensional system), here we discuss the Bell CGLMP parameter for general $d$-dimensional system.
\subsection{Bell inequality for general dimensional system}
We will consider $d$-dimensional systems. Each measurement can have $d$ possible end results: $A_1, A_2, B_1, B_2 =0,\ldots, d-1$. The Bell CGLMP parameter for general $d$-dimensional system is defined as
\begin{eqnarray}
I_d &=& \sum_{k=0}^{[d/2] -1} \left ( 1 - { 2 k \over d-1} \right) 
\\\nonumber
& & \Bigl( \Bigl[  P(A_1=B_1 + k) +  P(B_1= A_2+k +1) \\\nonumber
& & + P(A_2=B_2 + k ) +
  P(B_2=A_1+ k)\Bigr]\nonumber\\\nonumber
& & -\Bigl[  P(A_1=B_1 - k-1) +  P(B_1= A_2-k) \\\nonumber
& &+ P(A_2=B_2 - k-1 ) +
  P(B_2=A_1- k-1)\Bigr ] \Bigr)
\ .
\label{Id}
\end{eqnarray}
where we have defined the probability $P\left(A_a=B_b+\right.$ $k)$ as\cite{foot3},\cite{foot4}
\be\label{eq:PA=B+k}
P\left(A_a=B_b+k\right) \equiv \sum_{j=0}^{d-1} P\left(A_a=j+k \bmod d, B_b=j\right) .
\ee

We consider quantum states of $d$-dimensional entangled systems
\begin{equation}
|\psi\rangle = \frac{1}{\sqrt{\sum_{m,n=0}^{d-1} |\m_{m,n}|^2}} \sum_{m,n=0}^{d-1} \m_{m,n}|m\rangle_A \otimes
|n\rangle_B \ ,
\label{psi}
\end{equation} 
where we have normalised such that $\langle \psi|\psi\rangle=1$.
We will assume \cite{Belld} that the operators $A_a, a=1,2$ and $B_b, b=1,2$, measured by Alice and Bob respectively have the non-degenerate eigenvectors
\be
\begin{aligned}
|k\rangle_{A, a} &=\frac{1}{\sqrt{d}} \sum_{j=0}^{d-1} X^{(a)}_{j,k}|j\rangle_A, \\
|l\rangle_{B, b} &=\frac{1}{\sqrt{d}} \sum_{j=0}^{d-1} Y^{(b)}_{j,l}|j\rangle_B,
\end{aligned}
\ee
with $X^{(a)}_{j,k}=\exp \left(i \frac{2 \pi}{d} j\left(k+\alpha_a\right)\right), Y^{(b)}_{j,l}=\exp \left(i \frac{2 \pi}{d} j\left(-l+\beta_b\right)\right)$ and where $\alpha_1=0, \alpha_2=1 / 2, \beta_1=1 / 4$, and $\beta_2=-1 / 4$. As explained in \cite{Belld}, the measurements correspond to giving the states $|j\rangle_A$ and $|j\rangle_B$ a phase which depends on the choice of measurement followed by a measurement in the Fourier transform basis or an inverse Fourier transform basis. For the purpose of our paper, we will be content with using the resulting Bell parameter $I_3$ as a kind of ``order parameter'' to distinguish theories in the S-matrix space. Above, we have quoted the the measurement settings given in \cite{Belld}, which is what we will use. A potential generalization of this analysis is to replace $\alpha_a, \beta_b$ by $\alpha_{aj},\beta_{b j}$ and maximize $I_3$ over this 12 parameter space for a given state $|\psi\rangle$. General choice of $\alpha_{aj},\beta_{b j}$ will be discussed in appendix. 
In the set-up used in \cite{Belld} and summarized above, we obtain
\be\label{eq:PAkPBl}
\begin{split}
&P(A_a=k,B_b=l)\\
=&\langle \psi|\Big(|k\rangle_{A, a}\otimes |l\rangle_{B, b} ~~~{} _{A, a}\langle l| \otimes {}_{B, b} \langle k|\Big)|\psi\rangle\\
=& N \sum_{m,n,m',n'=0}^{d-1} \mu_{m',n'} \mu^*_{m,n}  X^{(a)}_{m,k} X^*_{m',k}Y^{(b)}_{n,l}Y^*_{n',l}\\
=& N  \left| \sum_{m,n=0}^{d-1} \m_{m,n}   X^{(a)}_{m,k} Y^{(b)}_{n,l}\right|^2 \,.
\end{split}
\ee
with $N=\frac{1}{d^2 \sum_{m,n=0}^{d-1} |\m_{m,n}|^2}$. We can put the above probability in eq \eqref{eq:PA=B+k}, and eq \eqref{I3} will give $I_3$ for generic $\m_{m,n}$. 

\subsection{Bell inequality in 2-2 pion scattering}

The pion transition amplitude can be written as \cite{ABPHAS}
\be\label{Mijdef}
\begin{split}
&\mathcal{M}_{a_1 a_2}^{a_3 a_4}(s, t, u)(s, t, u)= A(s \mid t, u) \delta_{a_1 a_2} \delta^{a_3 a_4}\\
&+A(t \mid u, s) \delta_{a_1}^{\ a_3} \delta_{a_2}^{\ a_4}+A(u \mid s, t) \delta_{a_1}^{\ a_4} \delta_{a_2}^{\ a_3}\,,
\end{split}
\ee
The index $a_i$ runs over $0,1,2$. Crossing symmetry on $A(s \mid t, u)$ is given by $A(s \mid t, u)=A(s \mid u, t)$. 
Now the external states are  isospin eigenstates. In nature, pions appear in mass eigenstates defined via
\be
|\pi^0\rangle=|0\rangle\,,~~|\pi^+\rangle=\frac{|1\rangle+i |2\rangle}{\sqrt{2}}\,,~~|\pi^-\rangle=\frac{|1\rangle-i |2\rangle}{\sqrt{2}}\,.
\ee
There are six independent in-states, namely $\pi^0 \pi^0,  \pi^+\pi^0, \pi^-\pi^0, \pi^+\pi^-, \pi^+\pi^+, \pi^-\pi^- $.

We will consider a generic final states, a generic superposition in isospin basis. Hence we find six different Bell expressions with different $\m^{m,n}_{\pi^0 \pi^0}, \m^{m,n}_{\pi^+\pi^0}, \m^{m,n}_{\pi^- \pi^0}, \m^{m,n}_{\pi^+\pi^-}, \m^{m,n}_{\pi^+\pi^+}, \m^{m,n}_{\pi^-\pi^-}$. For example
 $\m^{m,n}_{\pi^+\pi^-}$ can be figured out in the following manner
\be
\begin{split}
\m^{m,n}_{\pi^+\pi^-}&=\langle m|\otimes \langle n| M |\pi^+\rangle\otimes |\pi^-\rangle\\
&= \frac{\mM_{1,1}^{m,n}+\mM_{2,2}^{m,n}-i\mM_{1,2}^{m,n}+i \mM_{2,1}^{m,n}}{2}\,.
\end{split}
\ee
Similarly one can find out all other $\m$'s. Hence using these six set of $\m_{\pi \pi}$, we can easily find $I_3^{\pi \pi}$ using \eqref{eq:PA=B+k}, \eqref{eq:PAkPBl} and \eqref{I3}. We find that $$I_3^{\pi^+ \pi^0}=I_3^{\pi^- \pi^0}=0, ~~I_3^{\pi^+ \pi^+}=I_3^{\pi^- \pi^-}=-\frac{1}{\sqrt{3}}\,.$$
The non-trivial two are $I_3^{\pi^+ \pi^-}$ and $I_3^{\pi^0 \pi^0}$.

\begin{small}
\be
\begin{split}
&I_3^{\pi^+ \pi^-}=\frac{1}{9 \left(2\left| M_0\right|^2+3 \left| M_1\right|^2+\left| M_2\right|^2\right)}\Big[8 \left(2 \sqrt{3}+3\right) \left| M_0\right|^2\\
&-2 \left(\sqrt{3}+6\right) \left| M_2\right|^2+2 \left(\sqrt{3}-3\right) \left(M_0M_2^*+M_2 M_0^*\right)\Big]\\
&I_3^{\pi^0 \pi^0}=\frac{1}{9\left( \left| M_0\right|^2+2\left| M_2\right|^2\right)}\Big[4 \left(2 \sqrt{3}+3\right)\left|M_0\right|^2\\
&-4 \left(\sqrt{3}+6\right) \left| M_2\right|^2-2 \left(\sqrt{3}-3\right) \left(M_0 M_2^*+M_2 M_0^*\right)\Big]\,,
\end{split}
\ee
\end{small}
where $M_0=3 A_s+A_t+A_u,~M_1=A_t-A_u, ~M_2=A_t+A_u$ and we have shortened $A(s \mid t, u)=A_s$ and so on.

We can write a low energy expansion of $A(s \mid t, u)$ 
\be 
A(s \mid t, u)=\frac{s}{f_{\pi }^2}+\frac{b_3 s^2+b_4 (t-u)^2}{f_{\pi }^4}+\mathcal{O}(s^3)+\text{Loops}
\ee
For a finite angle $0\leq \theta\leq \pi$ and around $s\approx 4$, we get
\be
\begin{split}
&I_3^{\pi^+ \pi^-}= 2.87293-0.75138\frac{b_4}{f_{\pi }^2}+O\left(\frac{1}{f_{\pi }^3}\right)\,,\\
&I_3^{\pi^0 \pi^0}=2.87293+1.50275\frac{b_4}{f_{\pi }^2}+O\left(\frac{1}{f_{\pi }^3}\right)\,.
\end{split}
\ee
For $I_3^{\pi^+ \pi^-}$ when Bell inequality is violated, we have  $0\leq \frac{b_4}{f_\pi^2}\lesssim 1.16$, while for Bell violation of $I_3^{\pi^0 \pi^0}$, we must have $0\geq \frac{b_4}{f_\pi^2}\gtrsim-0.58$. The positivity bound is $0<\frac{b_4}{f_\pi^2} \lesssim 0.41$, see \cite{rastellipion,AZ, Fernandez:2022kzi}. In order to satisfy the positivity bounds, there will be Bell violations at $s\approx 4$. For $\chi$PT the experimental value $\left(\frac{b_4}{f_\pi^2}\right)_{\chi PT}=0.00967$, which will violate the Bell inequalities at $s \approx4$.

\section{S-matrix bootstrap: Distinguishing theories}

In this section we will consider the Bell parameter in pion scattering and photon scattering but in theory space generated by the S-matrix bootstrap. We have different expectations for the two cases. In the pion case, we expect resonances. The $\rho$-resonance was put in by hand and it can be expected that other resonances, for instance the $f_0$ resonance will emerge for a theory that approximates QCD. Near a resonance, there will be an enhancement of of the amplitude in such a way that the final state is approximately a direct product state.  Thus near the QCD point in theory space, we may expect a decrease in the Bell parameter. While this intuition bears out, we should point out that the correlation between the Bell parameter and other entanglement measures like concurrence, negativity and entanglement entropy is indirect---this we show in the appendix. For the photon case, at low energies we expect the same findings as we obtained in the EFT analysis. However, here we can probe further and ask up to what scale will the QED $f_2/g_2$ value serve as a boundary between Bell violating theories and non-Bell violating theories. We should expect that at least until other massive states like vector bosons become important in the analysis, the EFT findings should hold. 

\subsection{Pion scattering}\label{sec:pionscattering}
\begin{figure}[hbt!]
\centering
\begin{subfigure}{0.5\textwidth}
  \centering
  \includegraphics[width=0.8\linewidth]{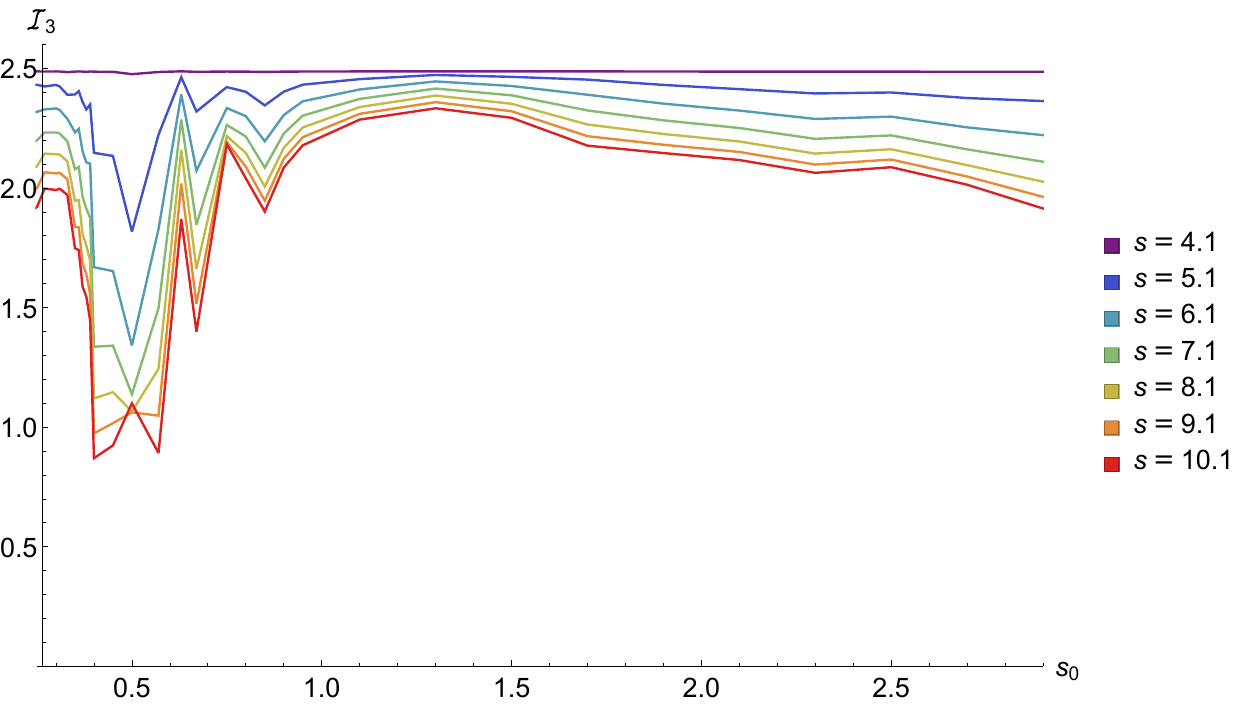}
  \caption{$I_3^{\pi^+ \pi^-}$ vs Adler zero $s_0$ of upper river S-matrices.}
  \label{fig:I3+-_4to11}
\end{subfigure}\\
\begin{subfigure}{.5\textwidth}
  \centering
  \includegraphics[width=0.8\linewidth]{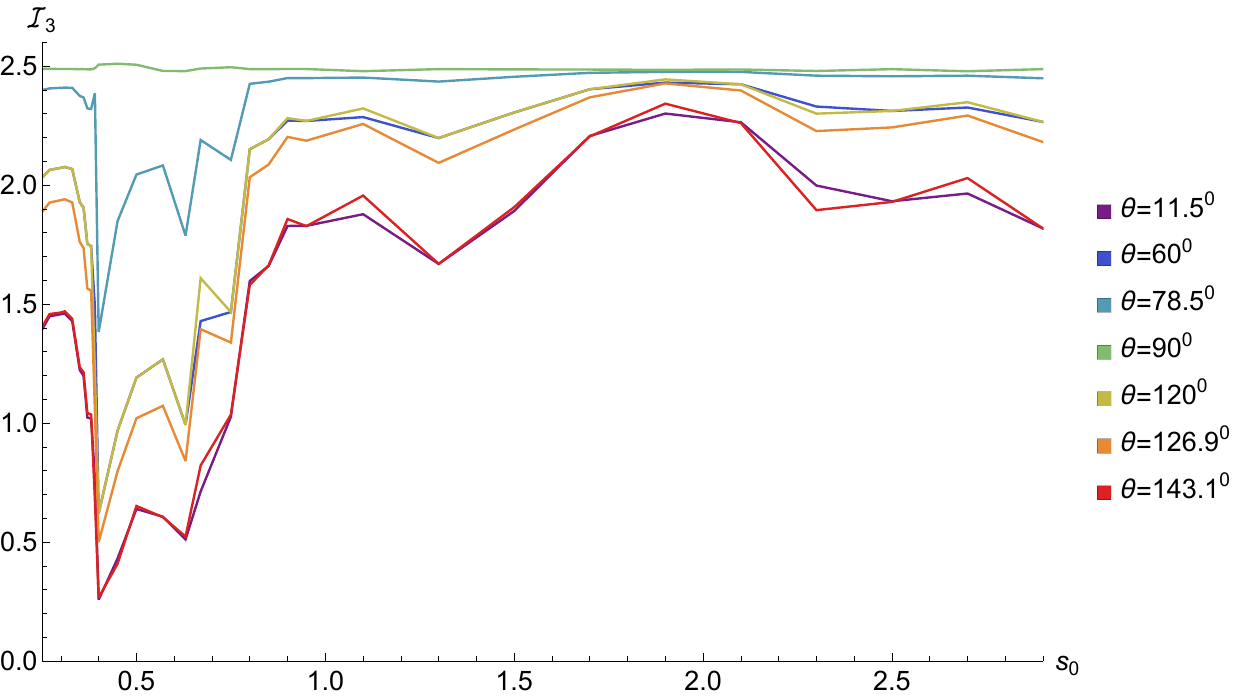}
 \caption{Averaged $\hat{I}_3^{\pi^+ \pi^-}$ vs Adler zero $s_0$ of upper river S-matrices for $s_{cut}=90$ for various $\theta$.}
\label{fig:I3+-_100to250}
\end{subfigure}
\caption{The $I_3^{\pi^+ \pi^-}$ and $I_3^{\pi^0 \pi^0}$ vs Adler zero $s_0$ for different values of $s, \theta, s_{max}$.}
\label{fig:I3}
\end{figure}

For  various S-matrix amplitudes in \cite{ABPHAS}, we can plot $I_3^{\pi^+ \pi^-}$ and $I_3^{\pi^0 \pi^0}$ vs Adler zero $s_0$ for different values of $s$. See figure \eqref{fig:I3} for details. The low energy behaviour of $I_3^{\pi^+ \pi^-}$ shows an interesting dip, which is exhibited for $s=4.1$ to $10.1$ in the forward limit, in figure \eqref{fig:I3+-_4to11}, while for high energy values we show in \eqref{fig:I3+-_100to250} for various values of angles $\theta$. We define an averaged $I_3$: 
\be\label{eq:avgI3} 
\hat{I}_3^{\pi^+ \pi^-}=\frac{1}{s_{max}-4}\int_{4}^{s_{max}}ds I_3^{\pi^+ \pi^-}\,,
\ee
which is presented in figure \eqref{fig:I3+-_100to250} with $s_{max}=90$. We find that the behaviour is same for various $s, \theta$ values form figure \eqref{fig:I3+-_4to11} and \eqref{fig:I3+-_100to250}. More interestingly both plots show a universal global minimum around $s_0\approx0.4$, which is very close to the QCD value $s_0=0.42$--see \cite{ABPHAS}. Further, the location of the minimum coincides with the S-matrices exhibiting Regge behaviour, i.e., the spin $J$ vs $m^2$ of the real part of the resonances lie approximately on a straight line \cite{ABPHAS}.

\subsection{Photon scattering}
\begin{figure}[hbt!]
\centering
\begin{subfigure}{0.5\textwidth}
  \centering
  \includegraphics[width=0.8\linewidth]{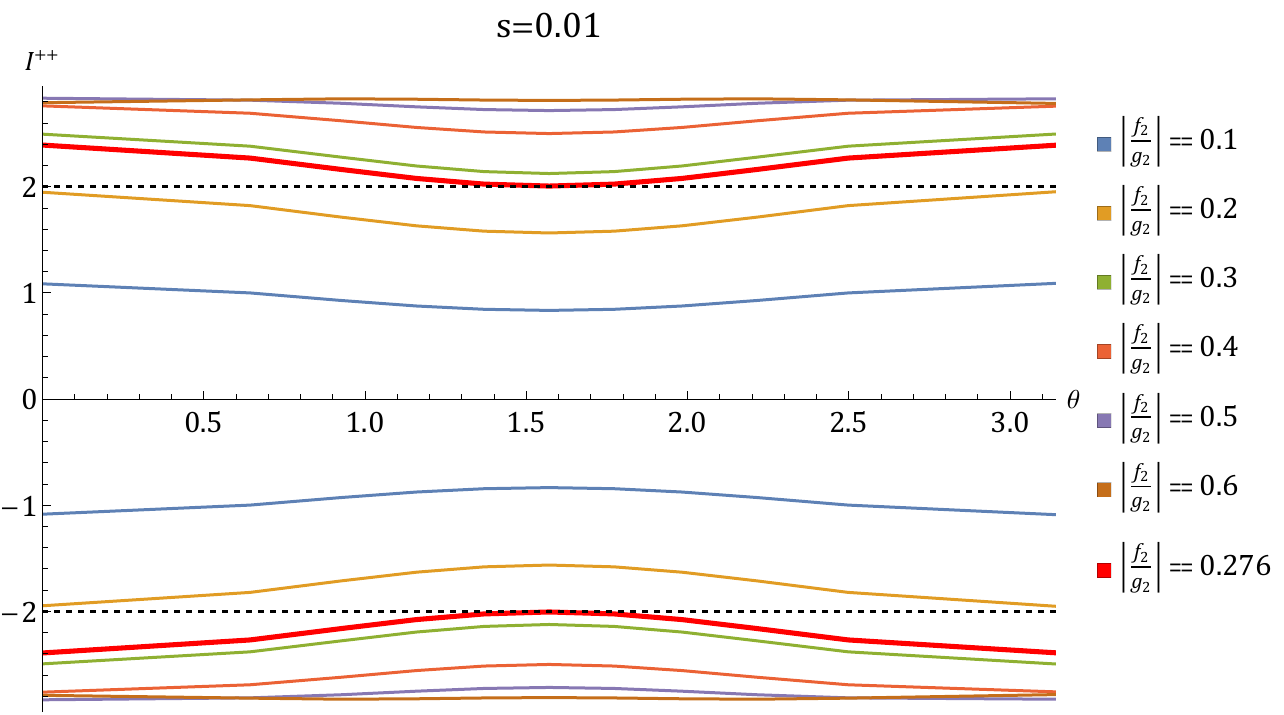}
  \caption{}
  \label{fig:photonlowschi}
\end{subfigure}\\
\begin{subfigure}{.5\textwidth}
  \centering
  \includegraphics[width=0.8\linewidth]{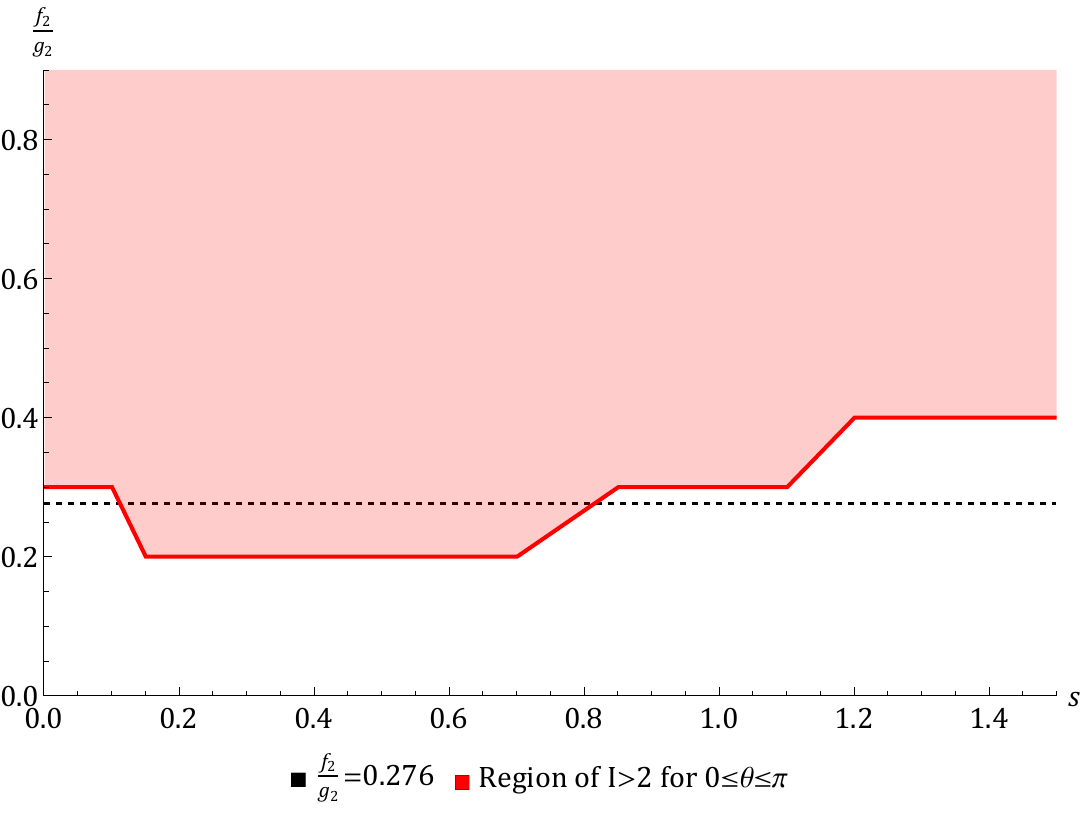}
 \caption{}
\label{fig:profilef2s}
\end{subfigure}
\caption{The behaviour of $I^{++}$ obtained from S-matrix bootstrap data \cite{joaophoton}, for various energy $s$, angle $\theta$ and $\bar{f}_2$. Figure (a): Low energy behaviour of $|I^{++}|$ vs $\theta$ for various $\bar{f}_2$. We choose low energy to be $s=0.01$. The Bell violation starts at $\left|\frac{f_2}{g_2}\right|>0.276$, which are indicated in red color. Figure (b): The profile of $f_2$ vs $s$ for which $I^{++}>2$ for all allowed values of $\theta$ using bootstrap data \cite{joaophoton}.}
\label{fig:I3photon}
\end{figure}

For  various S-matrix amplitudes in \cite{joaophoton}, we can plot $I^{++}=I^{--}$ vs $f_2/g_2$ for different values of $s$. Note that $I^{-+}=I^{+-}=0$ since $\phi_5\approx 0$ for these S-matrices. We plot the non-zero Bell parameter for the S-matrices obtained  in \cite{joaophoton}  by minimising\cite{foot5}  $\bar{g}_4$ vs $\bar{f}_2$ in \eqref{fig:I3photon}. Low energy behaviour of $|I^{++}|$ vs $\theta$ for various $\bar{f}_2$ is shown in figure \eqref{fig:photonlowschi}. We choose the energy\cite{foot6} to be $s=0.01$. In \cite{joaophoton}, discrete values of $f_2/g_2$ have been used. We use the $I_3$ for these values and find an interpolating function. We find that at low energy, Bell violation starts when $\left|\frac{f_2}{g_2}\right|>0.276$, which is the very close to the QED value $\left(\frac{f_2}{g_2}\right)_{QED}=-0.2736$. We see from the figure that our EFT analysis is in complete agreement. General profile of $f_2$ vs $s$ for which $I^{++}>2$ allowing all values of $\theta$ is shown in figure \eqref{fig:profilef2s}--in this figure, we have kept the $f_2/g_2$ values as the discrete choices used in \cite{joaophoton}. We see from figure \eqref{fig:profilef2s}, that to have minimum Bell violation for high energy one must have $|\bar{f}_2|\approx 0.3$, which is close to the QED value.

\section*{Discussion}
In this paper, we have investigated Bell violation in various scattering processes and have made very interesting observations. 

The most interesting one pertains to LbyL scattering.
For LbyL scattering at low energies, we have exhibited a novel fine-tuning problem in the Bell parameter, namely that the probability of detecting Bell violation using a continuous family of naturally motivated measurement settings at a randomly chosen scattering angle is very close to unity. Incorporating an axion or ALP removes this fine-tuning problem and predicts a line in the coupling--mass parameter plane on which the putative axion should lie. This prediction can be tested in near-future experiments that have already been designed to probe this parameter space.  A direct measurement of the Bell parameter in LbyL scattering at low energies seems prohibitively difficult. The smallness of the fine-structure constant and the fact that the process happens at one-loop makes the probability of scattering extremely tiny, being proportional to $\alpha^4 (s/m_e^2)^2$. As such, the simplistic set-up where one scatters photons off each other and sits with detectors off forward limit, gives time scales that are many times larger than the age of the universe. It will be interesting to re-visit this calculation in the context of heavy ion collisions \cite{atlas} where the energy scale is larger and hence the probability of scattering is bigger. However, this will be outside the purview of the effective field theory considerations used in this paper, where the energy scales considered are assumed to be lower than the hypothetical axion mass in the problem. Another point that needs further consideration is the inclusion of the graviton in the calculation. If the graviton fluctuates, then the notion of an angle becomes fuzzy and there should be some averaging over this fuzziness. We have not taken this into consideration in our analysis. We have simply assumed that quantum gravity effects should be suppressed compared to the axion physics. 

In the S-matrix bootstrap approach, it will be interesting to constrain the theory space by maximising the Bell parameter for the MES settings to see what the resulting S-matrices look like. One of the main motivations for us to examine Bell inequalities was to ask the question: How do we distinguish consistent theories from one another and why should a specific theory be the one that describes observed physics? To reiterate, we found nothing interesting in the entanglement measures such as entanglement entropy, negativity and concurrence which would distinguish the standard model from other theories. However, there was an almost definitive signature in the Bell parameter analysis, although we were not able to account for the sign. In the bootstrap, we assume Born rule to hold and typically impose unitarity and crossing symmetry as constraints. It will be interesting to revisit the analysis of this paper by relaxing some of these assumptions.

\section*{Acknowledgments} 
We thank Abhishek Dubey, Dipankar Home, Ranjan Laha, Alok Laddha, Arkajyoti Manna, Prashanth Raman, Barry Sanders, Urbasi Sinha and the participants of Bootstrapping Quantum Gravity $'23$ at KITP, Santa Barbara  for enlightening discussions. Special thanks to Ranjan Laha for comments on the draft and many discussions. AS acknowledges support DST, Govt. of India through the SERB core grant CRG/2021/000873. This research was supported in part by the National Science Foundation under Grant No. NSF PHY-1748958.

\onecolumngrid

{\begin{center}\bf \Large{Appendix}\end{center} }

\begin{appendix}

\section{The MES measurement parameters}\label{vimp}
\hskip 0.5cm \textbf{General measurement settings:}  We consider the measurement operators are of linear combinations of Pauli matrices as 
\be
\begin{array}{ll}
\hat{A}_1=\gamma_1 \mathbb{I}_2+\vec{C}_1 \cdot \vec{\sigma} &  \\
\hat{A}_2=\gamma_2 \mathbb{I}_2+\vec{C}_2 \cdot \vec{\sigma} & \\
\hat{B}_1=\delta_1 \mathbb{I}_2+\vec{D}_1 \cdot \vec{\sigma} &  \\
\hat{B}_2=\delta_2 \mathbb{I}_2+\vec{D}_2 \cdot \vec{\sigma} & 
\end{array}
\ee
where $\mathbb{I}_2$ is the $2\times 2$ identity matrix and $\sigma_1,~ \sigma_2,~\sigma_3$ are Pauli matrices. Restricting to eigenvalues of $\pm 1$, we set $\gamma_i=\delta_i=0$ and $\vert\vec{C}_i\vert=\vert\vec{D}_i\vert=1$. We parametrize $\vec{C}_a=\left\{ \sin\theta_a\cos\phi_a,~ \sin\theta_a\sin\phi_a,~ \cos\theta_a \right \}$ and $\vec{D}_a=\left\{ \sin\theta'_a\cos\phi'_a,~ \sin\theta'_a\sin\phi'_a,~ \cos\theta'_a \right \}$. This gives for the state $(|00\rangle+x|11\rangle)/\sqrt{1+x^2}$ (for notational simplicity, we denote the $-1,+1$ eigenstates by $|0\rangle,|1\rangle$ respectively) 
\be\label{eq:ranI}
\begin{split}
I=&\cos \theta _2 \left(\cos \theta '_2-\cos \theta '_1\right)+\cos \theta _1 \left(\cos \theta '_1+\cos \theta '_2\right)+\frac{2 x}{x^2+1}\\
&\times \Big[ \sin \theta _2 \cos \phi _2 \left(\sin \theta '_2 \cos \phi '_2-\sin \theta '_1 \cos \phi '_1\right)\\
&+\sin\theta_2\sin\phi_2\left(\sin\theta_1'\sin\phi_1'-\sin\theta_2'\sin\phi_2'\right)\\
&+
\sin \theta_1 \cos \phi _1 \left(\sin \theta '_1 \cos \phi '_1+\sin \theta '_2 \cos \phi '_2\right)\\
&-\sin\theta_1\sin\phi_1\left(\sin\theta'_1\sin\phi'_1+\sin\theta'_2\sin\phi'_2\right)\Big]
\end{split}
\ee
where $x=\frac{\Phi_1}{\Phi_2}$. 
Thus the general form of $I$ is
\be\label{genI3}
I=\g+\beta \frac{x}{x^2+1}\,.
\ee
If we demand maximum Bell violation for the maximally entangled state, $x=1$, we have $\g=2\sqrt{2}-\frac{\b}{2}$. 
%
 Further we must have $\beta\leq 4\sqrt{2}$ in order to avoid $ \vert I \vert >2\sqrt{2}$ for $x=-1$.  The minimum value of $\beta$ subject to the MES condition is $2\sqrt{2}$. This can be found using the explicit expression in terms of the angles given in \eqref{eq:ranI}.
Now we can dial $2\sqrt{2}\leq\b\leq 4\sqrt{2}$ and ask the range of $x$ for which $\vert I\vert >2$. 
For example when $\b=2\sqrt{2}$, we have $0.217<\vert x\vert <4.612$. The optimal bound on $x$ comes from $\beta=4\sqrt{2}$, which gives $\sqrt{2}-1\leq\vert x\vert \leq\sqrt{2}+1$.  Corresponding value of $I=\frac{4 \sqrt{2} x}{x^2+1}$. This is what we have found using our choice of measurement parameters, see eq (8) in the main text. 

\begin{figure}
  \centering
  \includegraphics[width=0.6\linewidth]{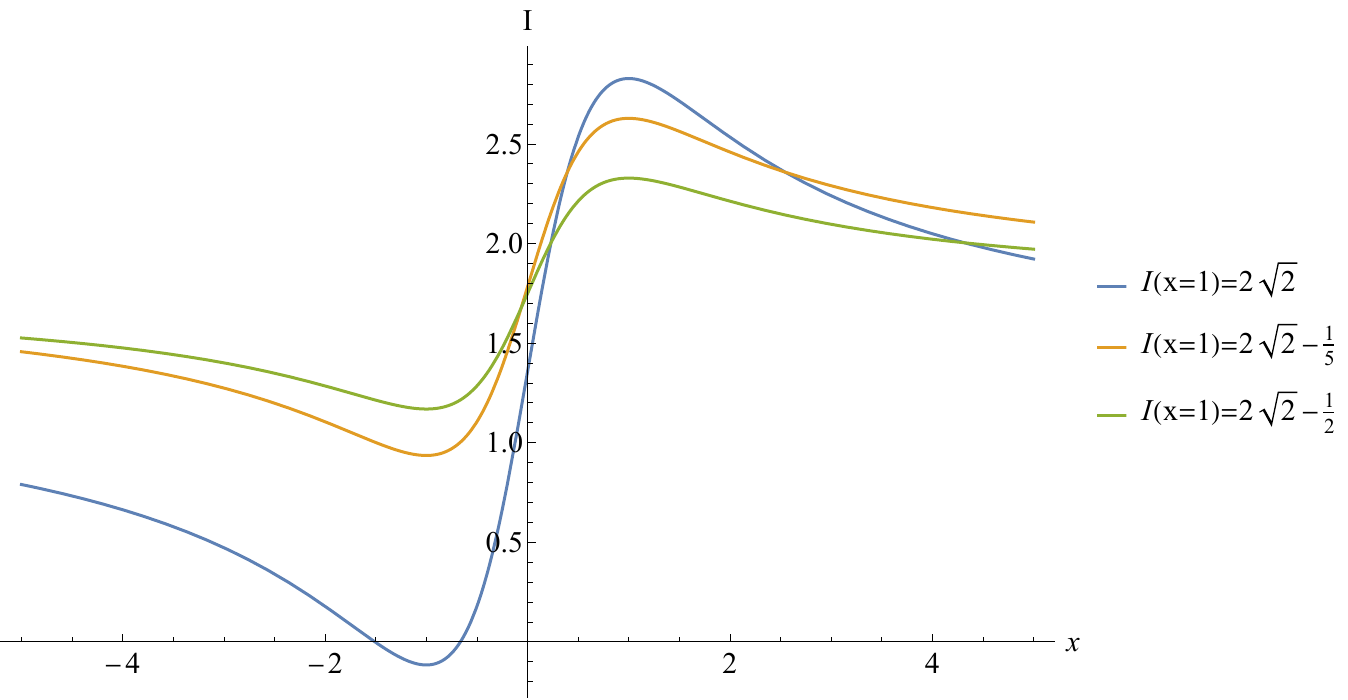}
 \caption{$I$  vs  $x$ for various $|I(x=1)|=2\sqrt{2}-\delta$, for $2\sqrt{2}-2>\delta>0$, then there is always some solution to $I_{max}(x)=2\sqrt{2}-\delta$ for $x\neq \pm 1$ for some settings and hence $I(x)$ will not be single-valued. If we don't fix the maximally entangled state to be $2\sqrt{2}-\delta$ then we can get multivalued answer.}
\label{fig:multival_I}
\end{figure}

\textbf{Why MES settings:} Why are we not considering all possible measurement settings? In order probe the theory space, we need $I(x)$ to be single-valued in $x$, to be able to distinguish theories. 
If we consider a state ${\mathcal N}(\Phi_1|00\rangle+\Phi_2|11\rangle)$ with $x=\Phi_1/\Phi_2$, then the maximum value of $I$ over {\it all measurement settings} can be shown to be $I_{max}(x)\equiv 2\sqrt{1+4x^2/(1+x^2)^2}\leq 2\sqrt{2}$, with equality only for $x=\pm 1$. The maximum Bell violation $I=2\sqrt{2}$ is clearly {\it \underline{only}} for the maximally entangled state. If for $x=1$ we have $|I|=2\sqrt{2}-\delta$, for $2\sqrt{2}-2>\delta>0$, then there is always some solution to $I_{max}(x)=2\sqrt{2}-\delta$ for $x\neq \pm 1$ for some settings and hence $I(x)$ will not be single-valued (see figure \eqref{fig:multival_I}). What can be checked is that if we restrict to settings with $|I(x=1)|$ fixed, then $I(x)$ is indeed single-valued. Furthermore, we would like maximum probability for Bell violation for SM. This is given by the MES settings.

\textbf{Details of simulations:}
In the main text, we have given the statistics for randomly choosing MES measurement parameters. We ask if we take some random MES measurement parameters, what is the probability of Bell violation in transverse direction. We use Mathematica to carry out the simulation. Note that \eqref{eq:ranI} has 8 variables. We can randomize all 8 variables and give $I$. From here we select those which give $I=2\sqrt{2}$ to some desired accuracy for $x=1$. We found this to be very time consuming. An easier way is the following. First randomize 4 variables in\eqref{eq:ranI} and for $x=1$ maximize with respect to the other 4 variables in \eqref{eq:ranI}. This way is significantly faster.

\begin{figure}[t]
\centering
\begin{subfigure}{0.5\textwidth}
  \centering
  \includegraphics[width=0.9\linewidth]{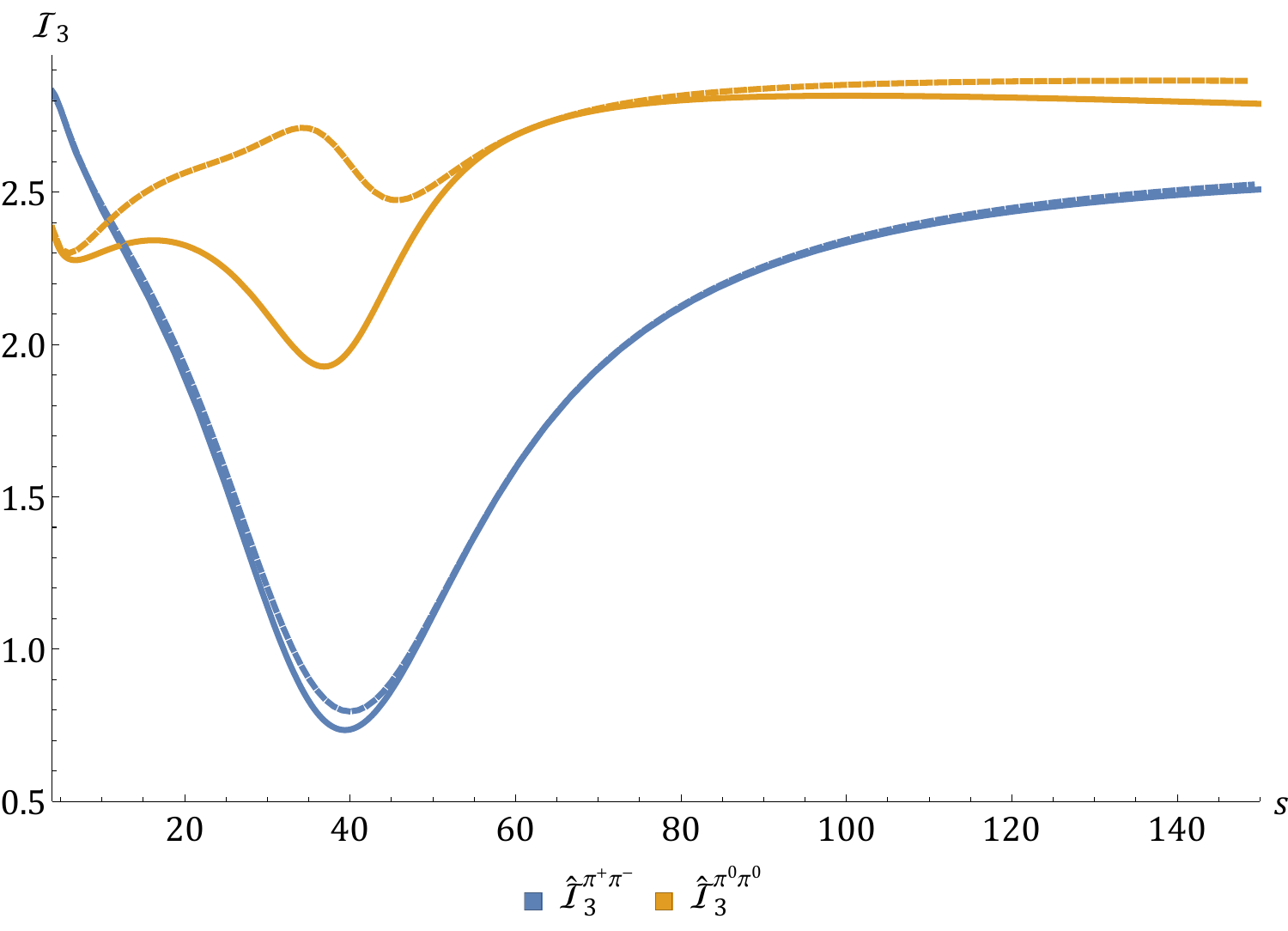}
 \caption{}
\label{fig:ChiPT_gen_measure}
\end{subfigure}%
\begin{subfigure}{.5\textwidth}
  \centering
  \includegraphics[width=0.9\linewidth]{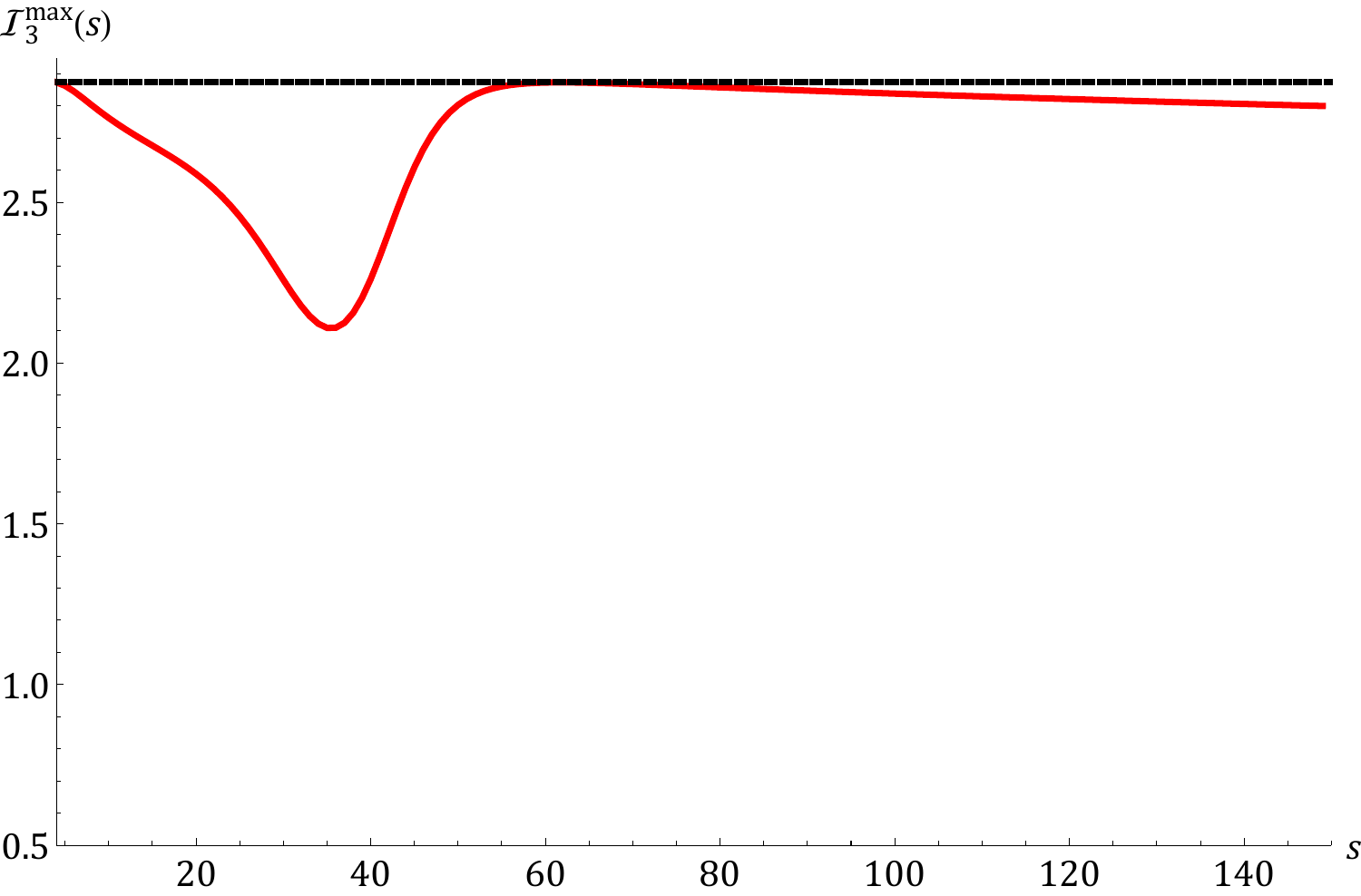}
 \caption{}
\label{fig:I3max}
\end{subfigure}\\
\caption{The choice of measurement parameters. (a) Solid lines are using  $\a_1=0,~\a_2=\frac{1}{2}, \beta_1=\frac{1}{4}, ~\beta_2=-\frac{1}{4}$. The dashed lines are the $I_3$ correspond to the maximising the $I_3$'s for each values of $s$.  (b) $I_3$ for maximally entangled state using the parameters which maximise $I_3^{\pi^0\pi^0}$ for each value of $s$. The dashed line is the value of $I_3=2.87293$ for maximally entangled state using $\a_1=0,~\a_2=\frac{1}{2}, \beta_1=\frac{1}{4}, ~\beta_2=-\frac{1}{4}$. }
\end{figure}

\textbf{Qutrit scenario}
Let us now address the qutrit scenario. 
Here we show an example which will explain why we choose our measurement parameters to be $\a_1=0,~\a_2=\frac{1}{2}, \beta_1=\frac{1}{4}, ~\beta_2=-\frac{1}{4}$. Let us consider the $I_3$'s for 2-loop $\chi$PT. Once we fix the parameters we get the $I_3$'s as a function of $s$. Here now we consider that $\a_1,~\a_2,\b_1~\b_2$ are arbitrary, then maximise the $I_3$s for each values of $s$, see figure \eqref{fig:ChiPT_gen_measure}. The dashed lines are the $I_3$ correspond to the maximising the $I_3$'s for each values of $s$. During this maximization process, we get a profile $\a_1(s),~ \a_2(s),~ \b_1(s), \b_2(s)$. Now we can compute the $I_3$ for maximally entangled state using this parameters, which is shown in figure \eqref{fig:I3max}. We see from the figure that all other parameter choice will always give lower (most cases) or equal (few points) values of $I_3$ for maximally entangled state than the choice of parameters $\a_1=0,~\a_2=\frac{1}{2}, \beta_1=\frac{1}{4}, ~\beta_2=-\frac{1}{4}$.

\section{Entanglement Entropy, Concurrence and Negativity}

\begin{figure}[b]
\centering
\begin{subfigure}{0.5\textwidth}
  \centering
  \includegraphics[width=0.8\linewidth]{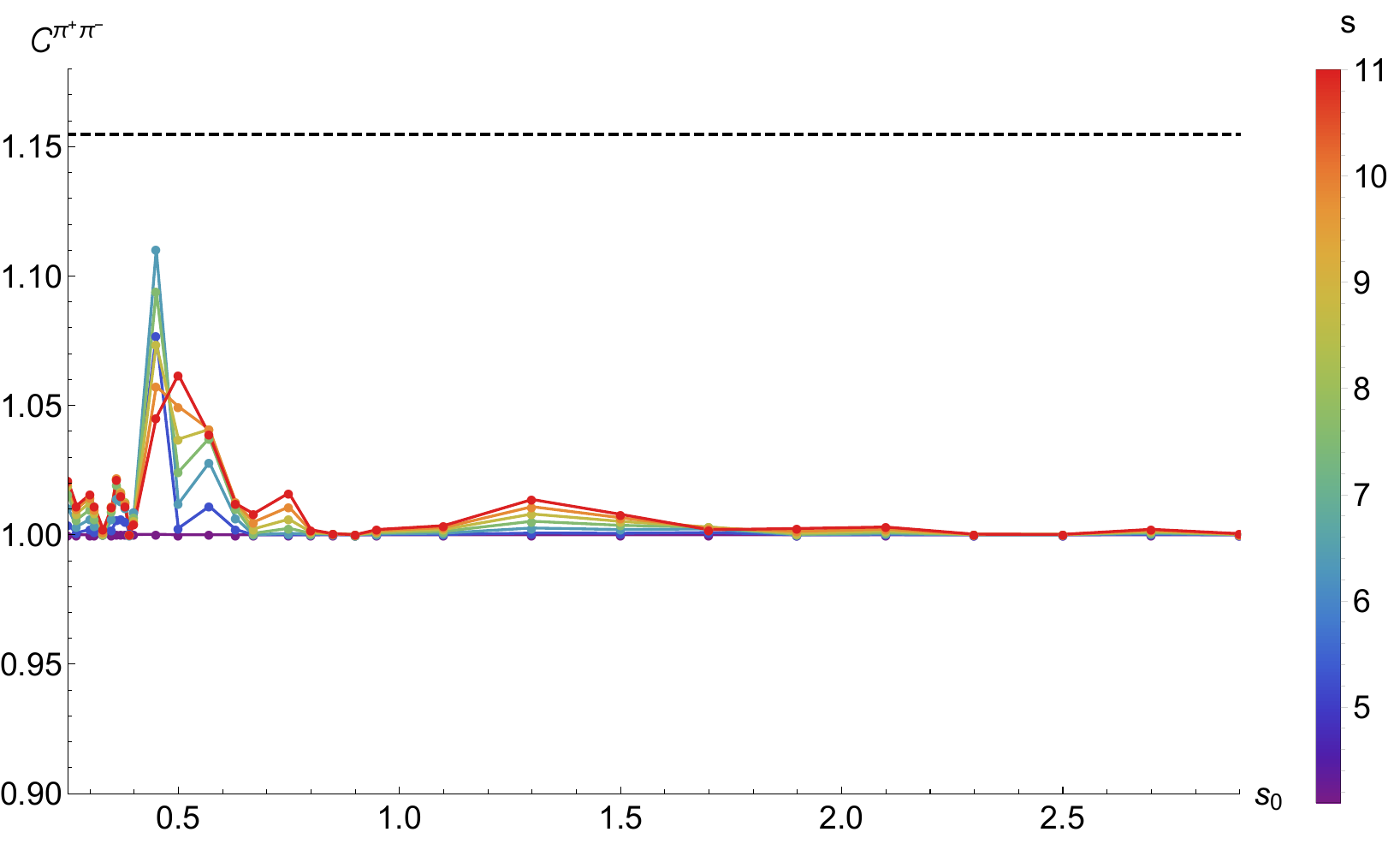}
  \caption{$\mathcal{C}^{\pi^+ \pi^-}$ vs $s_0$ of Upper River boundary}
  \label{fig:Cp}
\end{subfigure}%
\begin{subfigure}{.5\textwidth}
  \centering
  \includegraphics[width=0.8\linewidth]{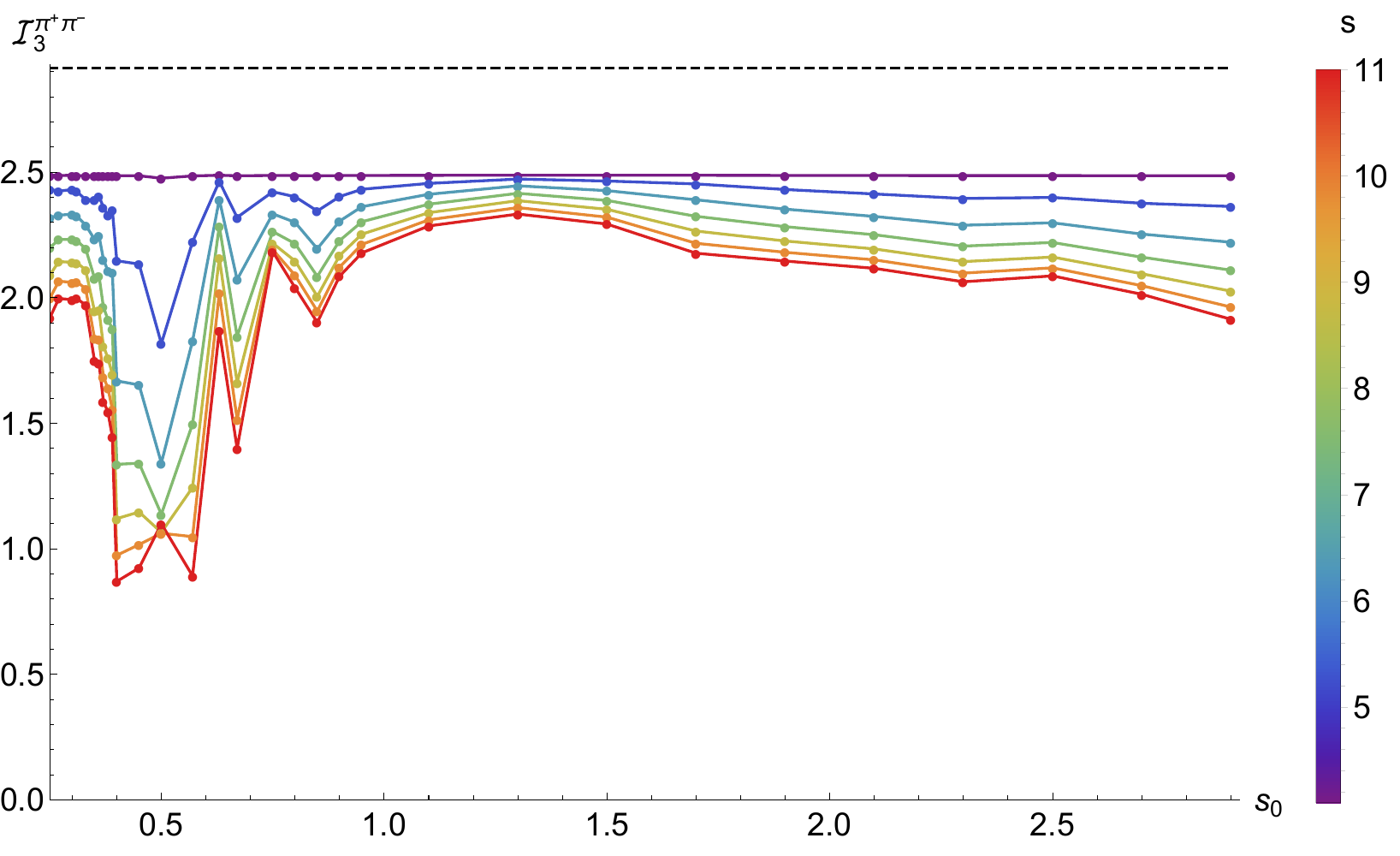}
 \caption{$I_3^{\pi^+ \pi^-}$ vs $s_0$ of Upper River boundary}
\label{fig:I3withCp}
\end{subfigure}\\
\begin{subfigure}{.5\textwidth}
  \centering
  \includegraphics[width=0.8\linewidth]{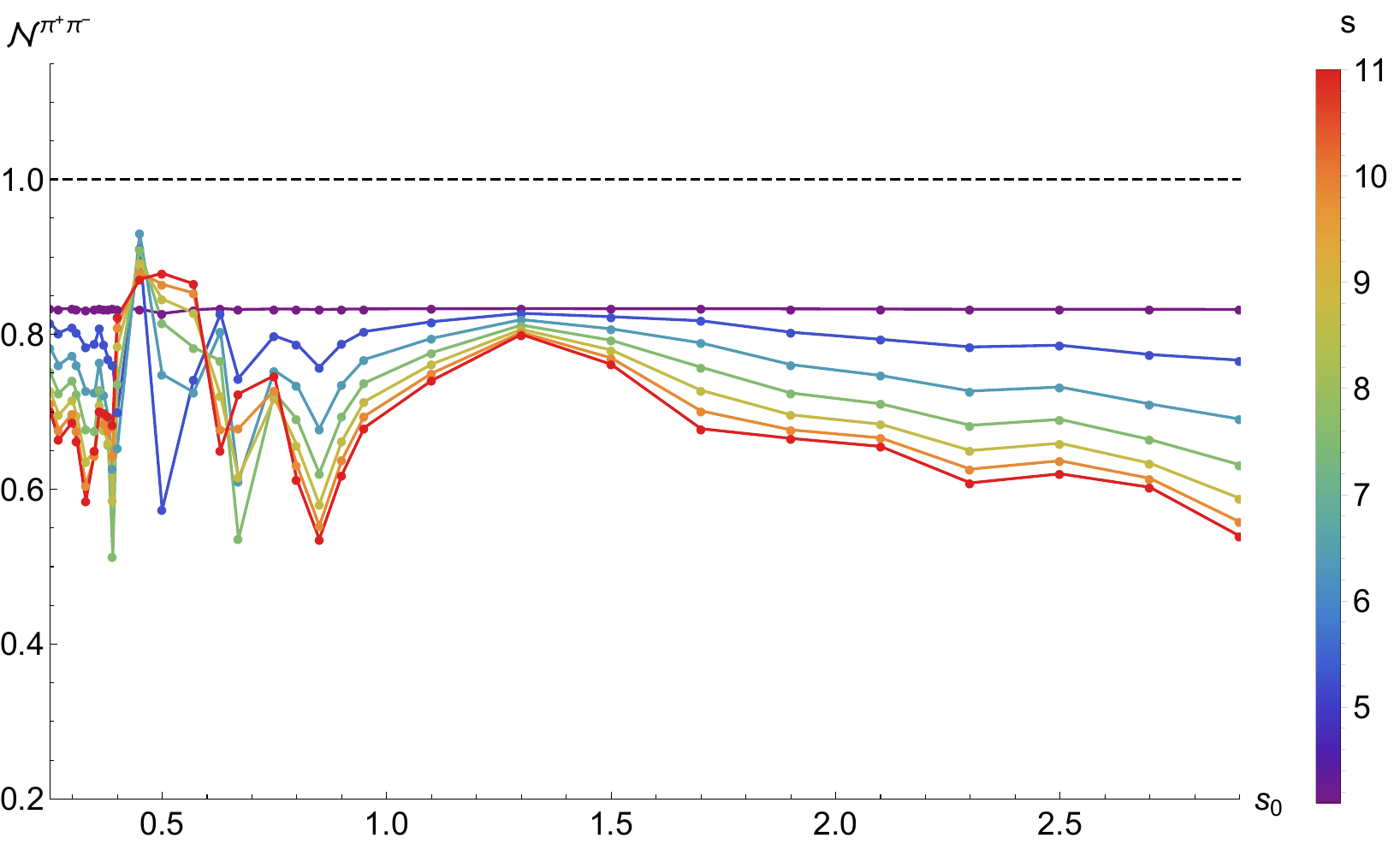}
  \caption{$\mathcal{N}^{\pi^+ \pi^-}$ vs $s_0$ of Upper River boundary}
  \label{fig:Np}
\end{subfigure}%
\begin{subfigure}{.5\textwidth}
  \centering
  \includegraphics[width=0.8\linewidth]{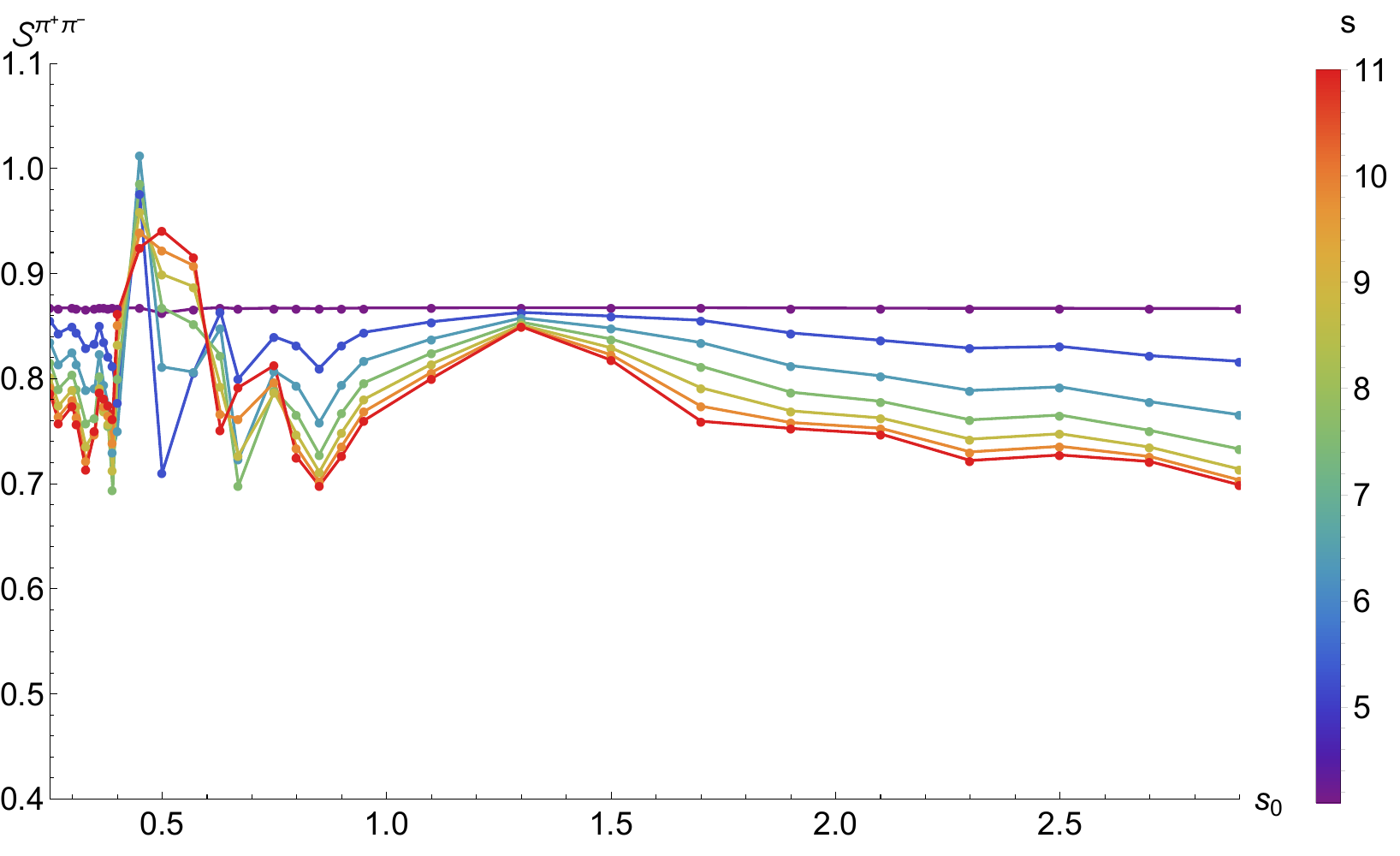}
 \caption{$\mathcal{S}^{\pi^+ \pi^-}$ vs $s_0$ of Upper River boundary}
  \label{ }
\end{subfigure}
\caption{$ \mathcal{C}^{\pi^+ \pi^-}, I_3^{\pi^+ \pi^-}, \mathcal{N}^{\pi^+ \pi^-}, \mathcal{S}^{\pi^+ \pi^-}$ vs the Adler zero $s_0$ labeling the S-matrices in \cite{ABPHAS}.}
\label{fig:SCNpion}
\end{figure}

In this section, we will compute entanglement entropy  $\mathcal{S}(\psi)$, concurrence $\mathcal{C}(\psi)$ and entanglement negativity $\mathcal{N}(\psi)$ for the d-dimensional system of Alice and Bob \cite{SCNPRA}.  We use Schmidt decomposition $|\psi\rangle=\sum \lambda_j\left|\alpha_j \beta_j\right\rangle$ of $\psi$, where the local Schmidt bases $\left|\alpha_j \beta_j\right\rangle$,
is obtained by local unitary transformations from local bases $|m\rangle_A \otimes
|n\rangle_B $. The formulas for $\mathcal{S}(\psi)$,  $\mathcal{C}(\psi)$ and $\mathcal{N}(\psi)$ are given by
\be
\begin{aligned}
\mathcal{S}(\psi)&=-\sum \l_j^2 \log(\l_j^2)\,,~~~
\mathcal{C}(\psi)=2 \sqrt{\sum_{j<k} \lambda_j^2 \lambda_k^2}\,,~~~~
\mathcal{N}(\psi)=\sum_{j<k} \lambda_j \lambda_k\,.
\end{aligned}
\ee

\subsection{For pions}
As we discussed in the main text that there are six mass eigenstates for two external pions, namely $|\psi\rangle_{\pi^0 \pi^0}, |\psi\rangle_{\pi^+\pi^0}, |\psi\rangle_{\pi^- \pi^0}, |\psi\rangle_{\pi^+ \pi^-}, |\psi\rangle_{\pi^+\pi^+}, |\psi\rangle_{\pi^-\pi^-}$. Out of theses six states, we focus on $|\psi\rangle_{\pi^0 \pi^0}, |\psi\rangle_{\pi^+ \pi^-}$ as discussed in section \eqref{sec:pionscattering}. One can compute $\m_{m,n}^{\pi^0 \pi^0}, \m_{m,n}^{\pi^+\pi^0}, \m_{m,n}^{\pi^- \pi^0}, \m_{m,n}^{\pi^+ \pi^-}, \m_{m,n}^{\pi^+\pi^+}, \m_{m,n}^{\pi^-\pi^-}$ using the prescription in section \eqref{sec:pionscattering}. Once we have these $\m_{m,n}$ we can compute the Schmidt decomposition and find $\l_{j}^{\pi^0 \pi^0}, \l_{j}^{\pi^+\pi^0}, \l_{j}^{\pi^- \pi^0}, \l_{j}^{\pi^+ \pi^-}, \l_{j}^{\pi^+\pi^+}, \l_{j}^{\pi^-\pi^-}$.  Figure \eqref{fig:SCNpion} shows the behaviour of $\mathcal{S}, \mathcal{C}, \mathcal{N}$ for various S-matrices that arise from the boostrap \cite{ABPHAS} as a function of $s$ in the forward limit. From these figures, we see that there is a correlation between where the dip in $I_3$ happens and the change in behaviour in $\mathcal{S}, \mathcal{C}, \mathcal{N}$. However, it is apparent that $\mathcal{N}$ and $\mathcal{S}$ have features in common while $\mathcal{C}$ behaves differently.

\subsection{For photons}

\begin{figure}[hbt]
\centering
\begin{subfigure}{0.5\textwidth}
  \centering
   \includegraphics[width=0.9\linewidth]{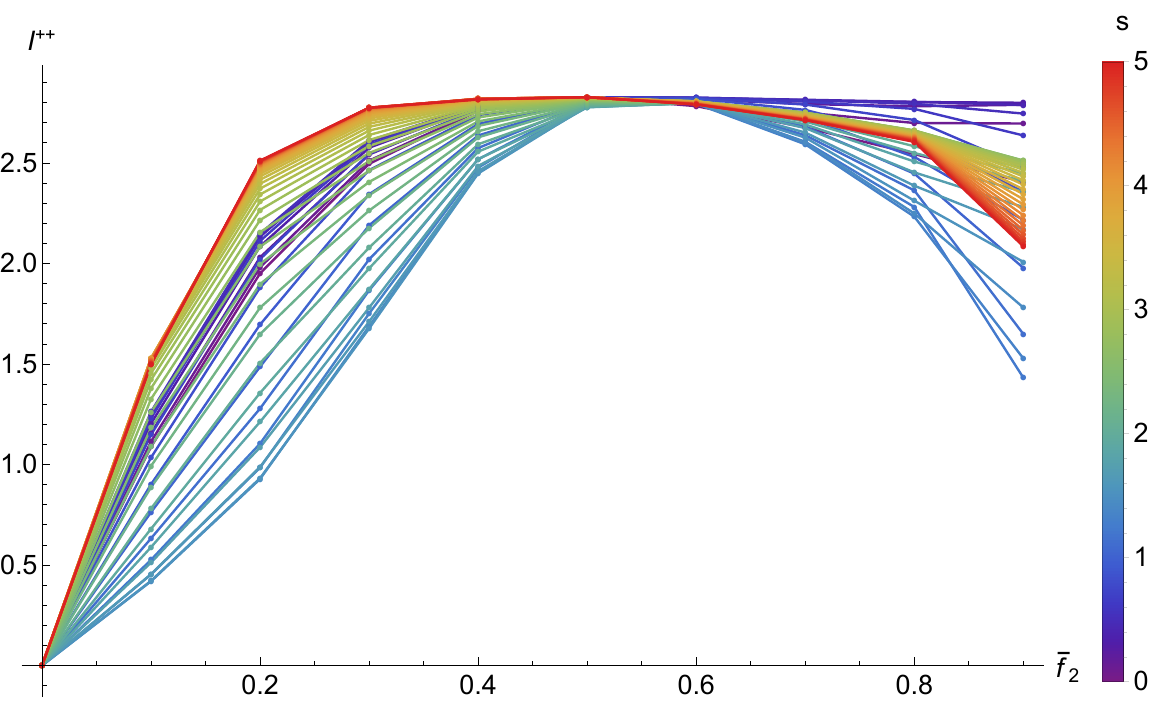}
 \caption{$I^{++}$ vs $\bar{f}_2$ for $s=0$ to $5$}
\label{fig:I3withC}
\end{subfigure}%
\begin{subfigure}{.5\textwidth}
  \centering
 \includegraphics[width=0.9\linewidth]{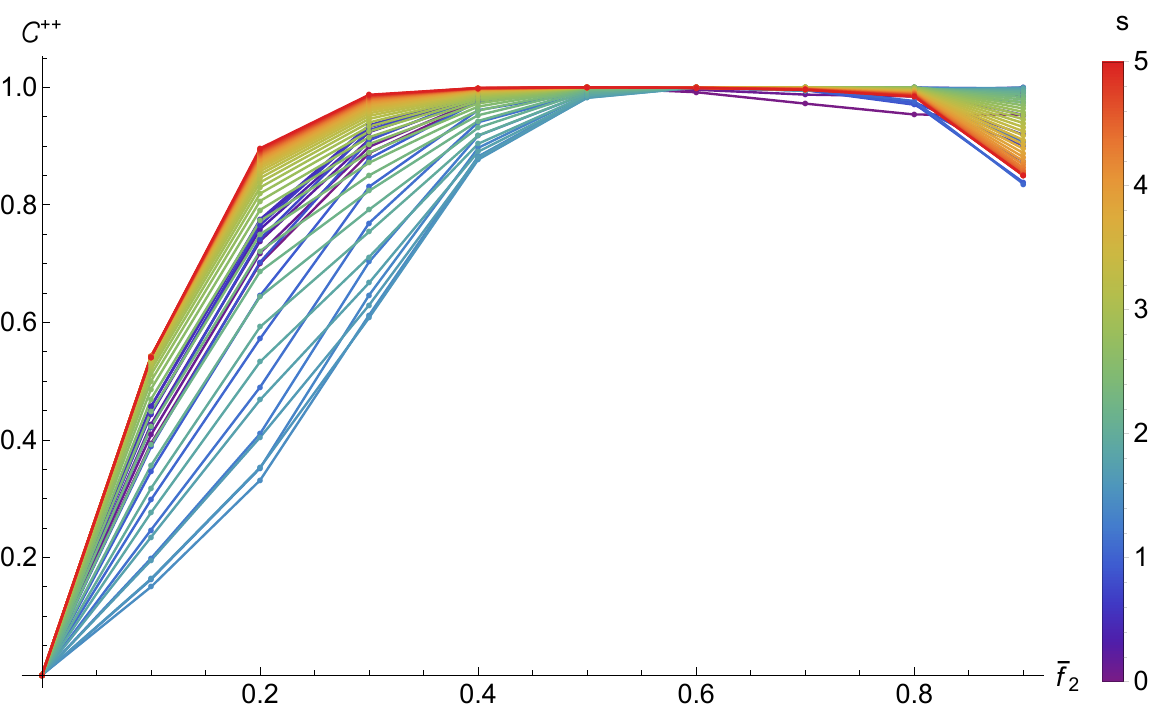}
  \caption{$\mathcal{C}^{++}$ vs $\bar{f}_2$ for $s=0$ to $5$}
  \label{fig:C}
\end{subfigure}\\
\begin{subfigure}{.5\textwidth}
  \centering
  \includegraphics[width=0.9\linewidth]{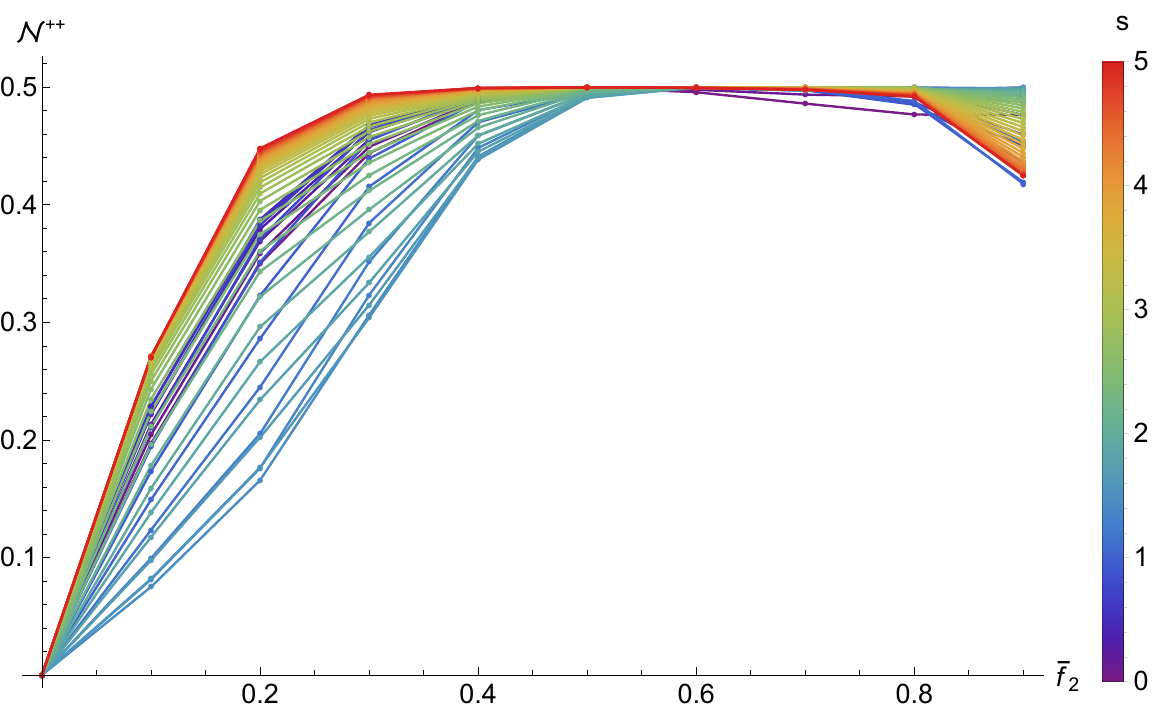}
  \caption{$\mathcal{N}^{++}$ vs $\bar{f}_2$ for $s=0$ to $5$}
  \label{fig:N}
\end{subfigure}%
\begin{subfigure}{.5\textwidth}
  \centering
  \includegraphics[width=0.9\linewidth]{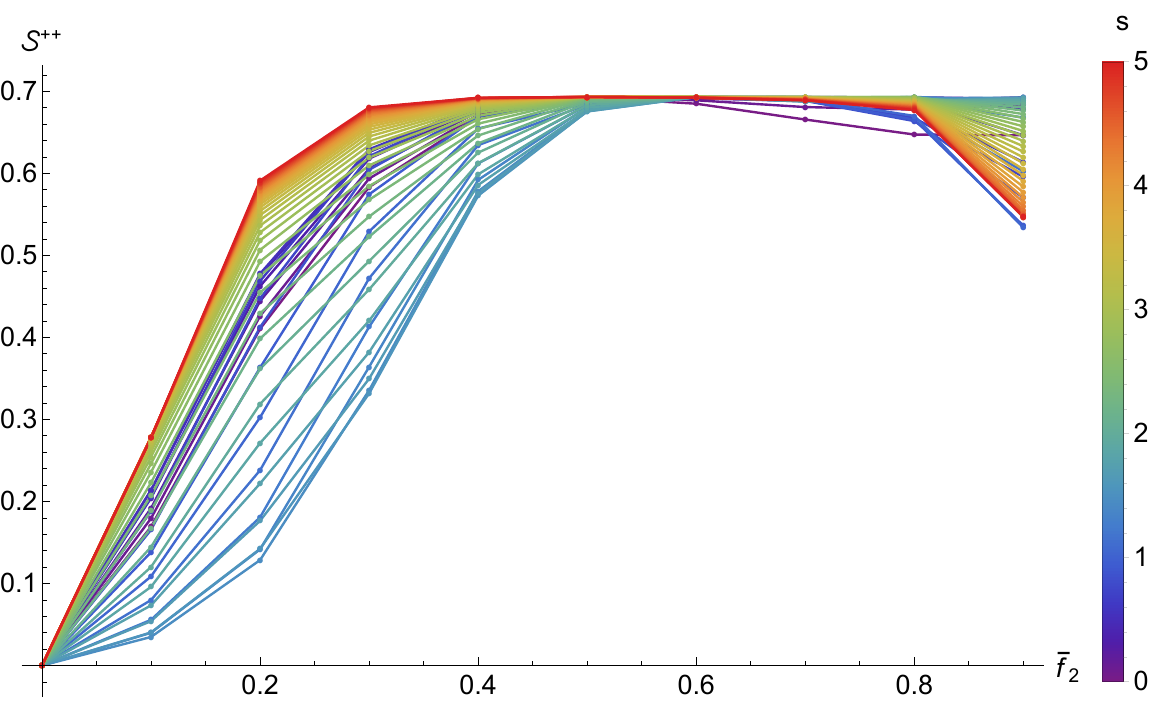}
 \caption{$\mathcal{S}^{++}$ vs $\bar{f}_2$ for $s=0$ to $5$}
  \label{fig:S}
\end{subfigure}
\caption{The $ I^{++}, \mathcal{C}^{++}, \mathcal{N}^{++}, \mathcal{S}^{++}$ vs $\bar{f}_2$ for various values of $s$.}
\label{fig:SCNphoton}
\end{figure}

We will compute the entanglement entropy, concurrence and entanglement negativity for photon amplitudes. They are given in figure \eqref{fig:SCNphoton}, which shows the behaviour of $\mathcal{S}, \mathcal{C}, \mathcal{N}$ for various S-matrices as a function of $s$ in the forward limit.
From the figures \eqref{fig:I3withC},  \eqref{fig:C}, \eqref{fig:S} and \eqref{fig:N} we see that $\mathcal{S}, \mathcal{C}, \mathcal{N}$ exactly mimic the behaviour of $I$, which is expected for Qubits.  In the plots we have used the S-matrices in \cite{joaophoton} which minimize  $\bar{g}_4$ vs $\bar{f}_2$.

\section{Bosonic string amplitudes and Bell inequalities}\label{ap:bosonic}
In this section we examine the tree level bosonic string amplitudes for graviton scattering. We want to examine how the Bell parameter behaviour changes due to the presence of stable resonances.

\begin{figure}[htb]
\centering
\begin{subfigure}{0.5\textwidth}
  \centering
  \includegraphics[width=0.9\linewidth]{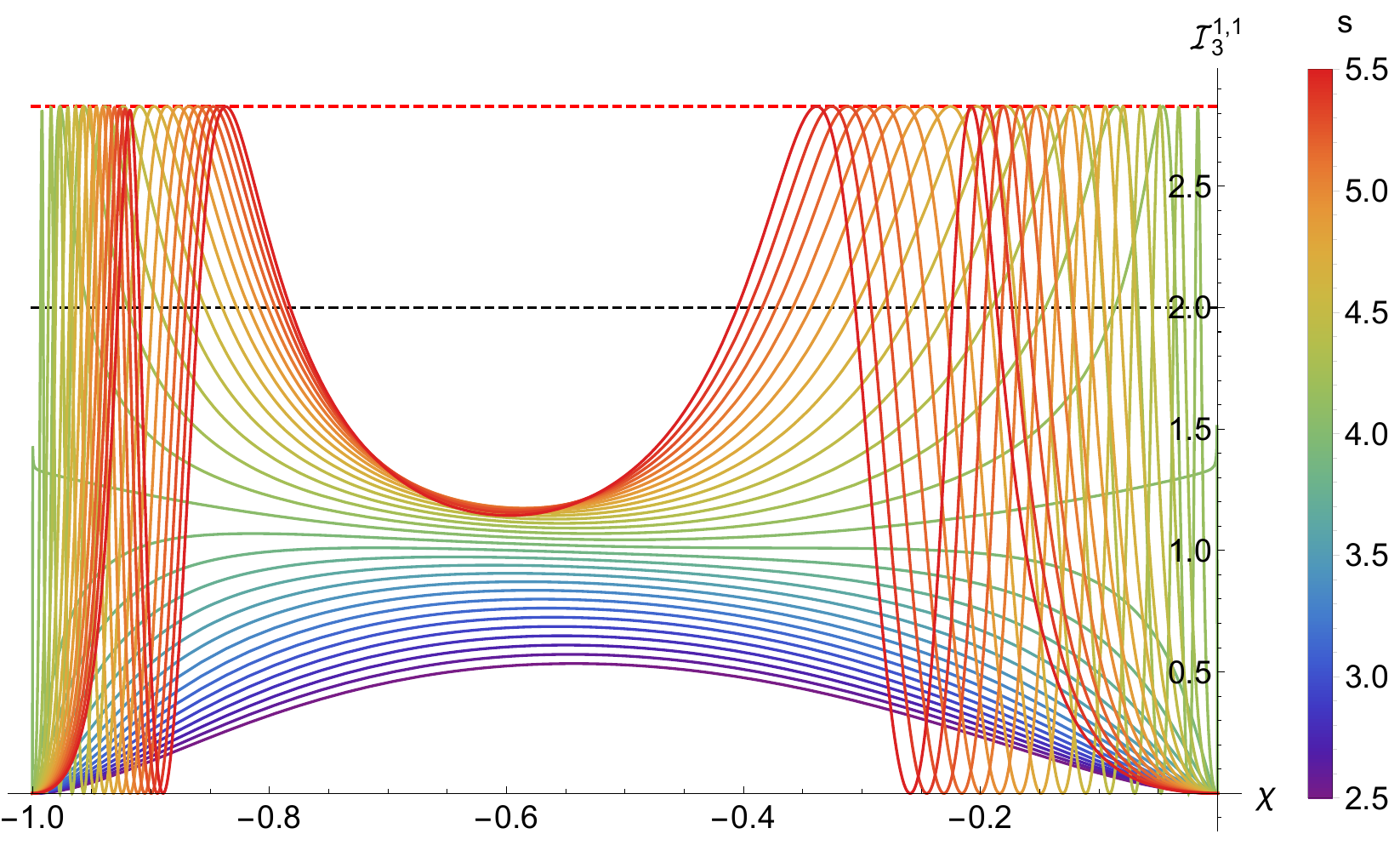}
  \caption{$I^{++}$ vs $\chi$ of bosonic graviton amplitudes for various $s$}
  \label{fig:bosongraviton11}
\end{subfigure}%
\begin{subfigure}{.5\textwidth}
  \centering
  \includegraphics[width=0.9\linewidth]{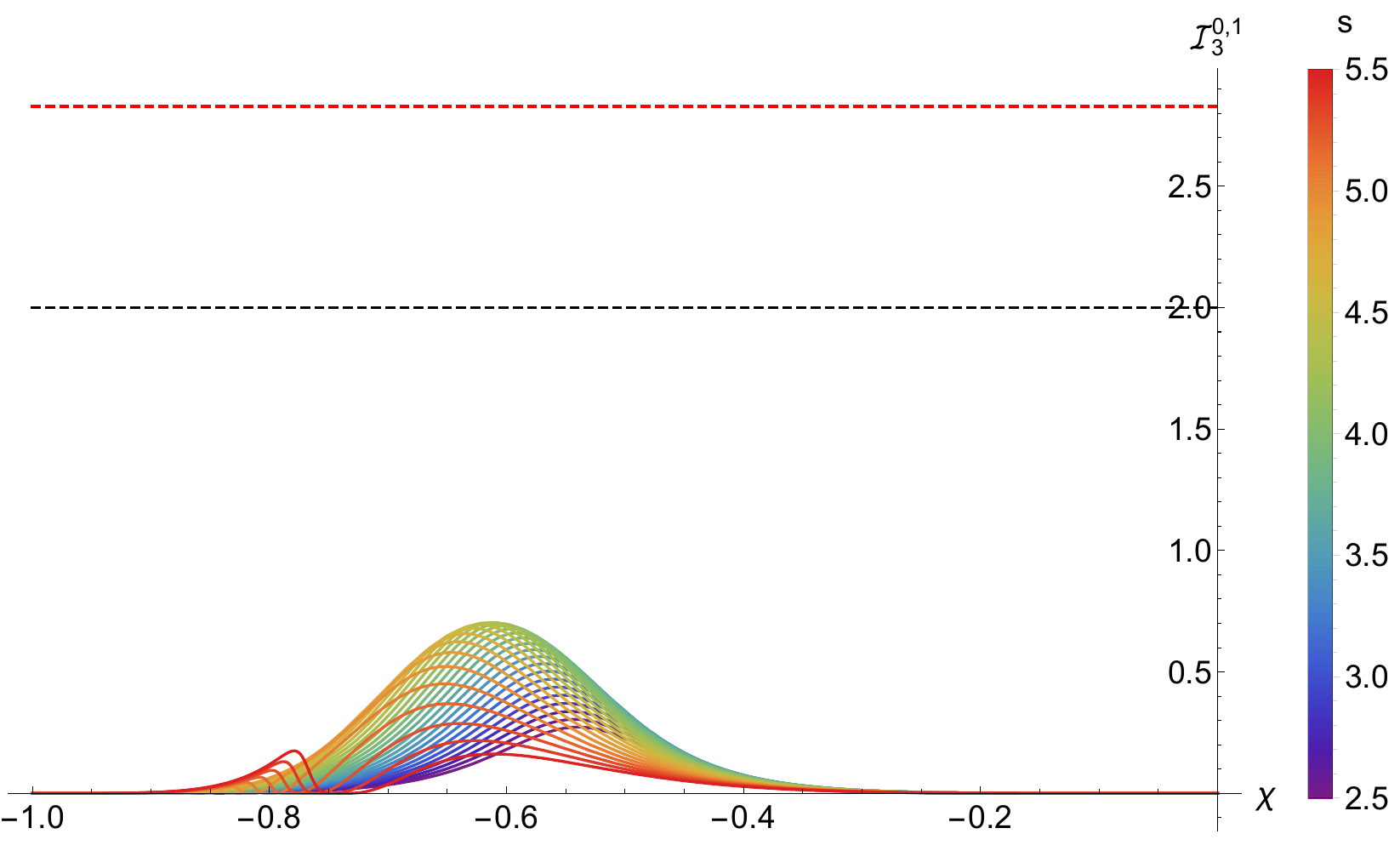}
 \caption{$I^{-+}$ vs $\chi$ of bosonic graviton amplitudes for various $s$}
\label{fig:bosongraviton01}
\end{subfigure}\\
\caption{Bell inequalities $I^{++}$  and $I^{-+}$ vs $\chi$ of tree level bosonic graviton amplitudes for various $s$}
\label{fig:bosongraviton}
\end{figure}

The 2-2 graviton scattering amplitudes are specified via:

\be
\begin{aligned}
\Phi_1(s, t, u) &=\frac{\alpha ^4 s^4}{256}  \left(1-\frac{\alpha ^2 s u}{16 \left(\frac{\alpha  t}{4}+1\right)}\right)^2 \frac{\sin \left(\frac{\pi  \alpha  t}{4}\right)\Gamma \left(-\frac{s \alpha}{4} \right)  \Gamma^2 \left(-\frac{t \alpha}{4} \right) \Gamma \left(-\frac{u \alpha}{4} \right)}{\Gamma \left(\frac{s \alpha }{4}+1\right) \Gamma \left(\frac{u \alpha }{4}+1\right)}\\
\Phi_2(s, t, u) &=\frac{\alpha ^6 s^2 t^2 u^2 }{4096}\left(1-\frac{1}{\frac{\alpha  s}{4}+1}-\frac{1}{\frac{\alpha  t}{4}+1}-\frac{1}{\frac{\alpha  u}{4}+1}\right)^2 \frac{\sin \left(\frac{\pi  \alpha  t}{4}\right)\Gamma \left(-\frac{s \alpha}{4} \right)  \Gamma^2 \left(-\frac{t \alpha}{4} \right) \Gamma \left(-\frac{u \alpha}{4} \right)}{\Gamma \left(\frac{s \alpha }{4}+1\right) \Gamma \left(\frac{u \alpha }{4}+1\right)}\\
\Phi_5(s, t, u) &=\frac{\alpha ^6 s^2 t^2 u^2 }{4096}  \frac{\sin \left(\frac{\pi  \alpha  t}{4}\right)\Gamma \left(-\frac{s \alpha}{4} \right)  \Gamma^2 \left(-\frac{t \alpha}{4} \right) \Gamma \left(-\frac{u \alpha}{4} \right)}{\Gamma \left(\frac{s \alpha }{4}+1\right) \Gamma \left(\frac{u \alpha }{4}+1\right)}
\end{aligned}
\ee
We find $I^{++}$ and $I^{-+}$ as shown in the figure \eqref{fig:bosongraviton}. We have considered $\a=1$.  We observe that at low energies, the Bell constraints are satisfied. As the resonances are approached, the Bell parameter begins to show characteristic dips and peaks indicative of the resonances.

\section{Gauss-Bonnet gravity in higher dimensions}
In this section, we briefly consider adding a Gauss-Bonnet term in higher dimensions ($d\geq 5$). We follow the notation in \cite{tolleynew}
\be
\begin{aligned}
\Phi_1(s, t, u) &=-\frac{8 (d-4) s^3 C_{\text{GB}}^2}{(d-2) \Lambda ^4}-18 \beta _3^2 s^3 \left(\frac{(d-4) s^2}{d-2}+2 s t+2 t^2\right)+\frac{s^3}{t u}\\
\Phi_2(s, t, u) &= \frac{12}{\Lambda ^4}\left( 5 \beta _3-\frac{2 (d-4) C_{\text{GB}}^2}{d-2}\right)~ s t u\\
\Phi_5(s, t, u) &=\frac{6 \beta _3}{\Lambda ^4} ~s t u
\end{aligned}
\ee
If we are interested in only $C_{GB}$, we can put $\beta_3=0$. We find that
\be
I^{++}=\frac{96 \sqrt{2} (d-4) \chi ^2 (\chi +1)^2 c_{\text{GB}}^2 \left(-8 (d-4) \chi  (\chi +1) c_{\text{GB}}^2-d+2\right)}{64 (d-4)^2 \chi ^2 (\chi +1)^2 \left(9 \chi ^2 (\chi +1)^2+1\right) c_{\text{GB}}^4+16 (d-4) (d-2) \chi  (\chi +1) c_{\text{GB}}^2+(d-2)^2}
\ee
where $c_{\text{GB}}=C_{\text{GB}} \frac{s}{\Lambda^2}$. We find $I^{-+}=0$. For $-1\leq \chi=1+2t/s \leq 0$, since the Bell parameter vanishes at $\chi=0$ there is no choice of $c_{\text{GB}}$ where Bell violation occurs for all physical angles. For the range
\be
|c_{\text{GB}}|\lesssim 0.42\sqrt{\frac{d-2}{d-4}}~~\text{   or   }~~0.62\sqrt{\frac{d-2}{d-4}} \lesssim|c_{\text{GB}}|\lesssim 0.85\sqrt{\frac{d-2}{d-4}}
\ee
there is no Bell violation, while 
there will always be some violation for some values of $\chi$ for $c_{\text{GB}}$ outside this range. In $d=4$ for the $R^3$ case, we had found similar bounds. 

\section{Bhabha scattering amplitudes}\label{ap:bhabhaamps}
In the low energy limit $s\approx 4m^2$ and generic scattering angle $\chi=1+2t/(s-4m^2)$, we have
\be
\begin{split}
&\mM_{--}^{--}(s,t)=\mM_{--}^{++}(s,t)=\frac{8 e^2 m^2-\kappa ^2 m^4}{2 \chi (s-4m^2) }+O((s-4m^2)^0)\\
&\mM_{--}^{-+}(s,t)=\mM_{--}^{+-}(s,t)=\frac{i \left(8 e^2 (2 \chi +5)+3 \kappa ^2 m^2 (2 \chi +1)\right)}{16 \sqrt{\frac{\chi }{\chi +1}}}+O(s-4m^2)\,.
\end{split}
\ee
where $\kappa=\sqrt{8\pi G}$. For small scattering angle $\chi $, we have
\be
\begin{split}
&\mM_{--}^{--}(s,t)=\mM_{--}^{++}(s,t)=\frac{8 e^2 m^2-\kappa ^2 m^4}{2 \chi (s-4m^2) }+O(\chi^0)\\
&\mM_{--}^{-+}(s,t)=\frac{i \left(16 e^2 \left(3 m+\sqrt{s}\right)+\kappa ^2 m \left(6 m^2+2 m \sqrt{s}-s\right)\right)}{8 \sqrt{\chi } \left(2 m+\sqrt{s}\right)}+O(\chi^0)\\
&\mM_{--}^{+-}(s,t)=\frac{i \left(16 e^2 \left(m+\sqrt{s}\right)+\kappa ^2 m \left(2 m^2+2 m \sqrt{s}+s\right)\right)}{8 \sqrt{\chi } \left(2 m+\sqrt{s}\right)}+O(\chi^0)
\end{split}
\ee
Around the pole $s=M_z^2$, we have up to $O\left(\left(s-M_z^2\right){}^0\right)$
\be
\begin{split}
&\mM_{--}^{--}(s,t)=\frac{2 e^2 m \csc ^2\left(2 \theta _w\right) \left(m \left(-2 \cos \left(2 \theta _w\right)+\cos \left(4 \theta _w\right)-2 \chi  (\chi +1)+1\right)+\chi  (\chi +1) M_z\right)}{s-M_z^2}\\
&\mM_{--}^{++}(s,t)=\frac{2 e^2 m \csc ^2\left(2 \theta _w\right) \left(m \left(2 \left(\chi ^2+\chi +1\right)-2 \cos \left(2 \theta _w\right)+\cos \left(4 \theta _w\right)\right)-\chi  (\chi +1) M_z\right)}{s-M_z^2}\\
&\mM_{--}^{-+}(s,t)=\frac{i e^2 m \sqrt{\chi } \sqrt{\chi +1} \csc ^2\left(2 \theta _w\right) \left((2 \chi +1) \left(M_z-2 m\right)-\sqrt{M_z^2-4 m^2} \left(1-2 \cos \left(2 \theta _w\right)\right)\right)}{s-M_z^2}\\
&\mM_{--}^{+-}(s,t)=-\frac{i e^2 m \sqrt{\chi } \sqrt{\chi +1} \csc ^2\left(2 \theta _w\right) \left((2 \chi +1) \left(2 m-M_z\right)-\sqrt{M_z^2-4 m^2} \left(1-2 \cos \left(2 \theta _w\right)\right)\right)}{s-M_z^2}
\end{split}
\ee

\section{EFT Lagrangians}
For convenience, we present here the lagrangians and relate the parameters appearing therein to the parameters used in the amplitudes. More details can be found in \cite{vichi1, joaophoton, bern, chuotgrav}.
\subsection{Euler-Heisenberg Lagrangian:}

Keeping in mind the 2-2 scattering of photon, the form of Lagrangian density we consider,
\be
\mathcal{L}_{\mathrm{EFT}}=-\frac{1}{4} F_{\mu \nu} F^{\mu \nu}+\mathcal{L}_8+\mathcal{L}_{10}+ \ldots,
\ee
the mass dimension $n$  part $\mathcal{L}_n$  read as
\be
\begin{aligned}
\mathcal{L}_8  =c_1\left(F_{\mu \nu} F^{\nu \mu}\right)\left(F_{\alpha \beta} F^{\beta \alpha}\right)+c_2\left(F_{\mu \nu} F^{\nu \rho} F_{\rho \sigma} F^{\sigma \mu}\right)  \,,~~~
\mathcal{L}_{10}   =c_3\left(F_{\alpha \beta} \partial^\beta F_{\mu \nu} \partial^\alpha F^{\nu \rho} F_\rho{ }^\mu\right)+\dots  \\
\end{aligned}
\ee
The EFT is valid up to some cut-off $\Lambda$. In four space-time dimensions, we have
$$
F_{\mu \nu} F^{\nu \sigma} F_{\sigma \rho} F^{\rho \mu}=\frac{1}{4}\left(F^{\mu \nu} \tilde{F}_{\mu \nu}\right)^2+\frac{1}{2}\left(F^{\mu \nu} F_{\mu \nu}\right)^2,
$$
with $\tilde{F}_{\mu \nu}=\frac{1}{2} \epsilon^{\mu \nu \rho \sigma} F_{\rho \sigma}$. The coefficients $c_i$ are called Wilson coefficients, they are related to $g_2,f_2, h_3$ as 
\be
g_2=2(4c_1+3c_2)\,, \quad f_2=2(4c_1+c_2)\,, \quad h_3=\frac{3}{3}c_3
\ee
The Wilson coefficients are given in table \eqref{tab:photontheory} for various theories, $m$ being the mass of the particle integrated out while $\alpha$ is the coupling between this particle and photons in the original theory. Of the particles listed in the table, only the axion and parity-odd spin-2 can achieve saturation of the Bell violation. The spinor case leads to an overall common factor that drops out in $f_2/g_2$. We focus on the simplest and well-motivated axion case.

\begin{table}[h]
\centering
\begin{tabular}{| C | C | C |}
\hline
  & g_2 & f_2   \\
  \hline
  \textbf{Spin }0  & &  \\
  \hline
  \text{scalar} & \frac{\alpha^2}{m^4} & \frac{\alpha^2}{m^4} \\ 
  \text{axion} & \frac{\alpha^2}{m^4} & -\frac{\alpha^2}{m^4}  \\
  \hline
  \textbf{Spin }2 & &   \\ 
  \hline
  \text{parity even I}   & \frac{\alpha^2}{m^4} & \frac{\alpha^2}{m^4}  \\ 
 \text{ parity even II}  & \frac{\alpha^2}{m^4} & 0  \\ 
 \text{ parity odd} & \frac{\alpha^2}{m^4} & -\frac{\alpha^2}{m^4}   \\
  \hline
  \textbf{One loop} & &  \\ 
  \hline
  \text{scalar QED} & \frac{2\alpha^2}{45 m^4} & \frac{\alpha^2}{30 m^4}   \\ 
  \text{spinor QED} & \frac{11\alpha^2}{45 m^4} & -\frac{\alpha^2}{15 m^4} \\ 
  \text{vector QED} & \frac{14\alpha^2}{5 m^4} & \frac{\alpha^2}{10 m^4}  \\
 \hline
\end{tabular}
 \caption{Wilson coefficients \cite{vichi1, joaophoton}.}
 \label{tab:photontheory}
\end{table}

\subsection{Effective Lagrangian for gravity:}
$$
\begin{aligned}
S=\frac{1}{16 \pi G} \int d^4 x & \sqrt{-g}\left[R-\frac{\beta_3}{3 !}  R^{(3)} +\ldots\right]+S_{\text {matter }}
\end{aligned}
$$
where we defined
$
R^{(3)}=R_{\mu \nu}{ }^{\rho \sigma} R_{\rho \sigma}{ }^{\alpha \beta} R_{\alpha \beta}{ }^{\mu \nu}.
$

\end{appendix}


\begin{thebibliography}{99}
\bibitem{foot1}
Since \cite{joaophoton} provides the data at discrete values of $f_2/g_2$, we are unable to do a fine-grained analysis using their data.
\bibitem{PAB}
The reason for this definition is understood by considering $P(A_1=B_1)$. Obviously this is $P(A_1=0, B_1=0)+P(A_1=1, B_1=1)$.
\bibitem{wcom} Vector bosons contribute with an opposite sign (see table-1 in appendix) and would have any way pushed $I$ in the opposite direction.
\bibitem{spin2oddv2}

Instead of ALP if we have spin 2 parity odd particle of mass $m_X$, the bound on the coupling  $g_X \gtrsim \frac{7.5\times 10^{-5}}{m_e^2 m_X}$, where definition of the coupling can be read of from the Lagrangian contribution $\mathcal{L}_X=\frac{g_X}{2}\left(h^{\mu \nu} \partial_\mu F_{\rho \sigma} \partial_\nu \tilde{F}^{\rho \sigma}-\frac{1}{4} \partial_\mu \partial_\nu h^{\mu \nu} F_{\rho \sigma} \tilde{F}^{\rho \sigma}\right)$.
\bibitem{foot2}
 Note: $I$=0 for tree level superstring theory for both photon and graviton scenarios \cite{nima}. For bosonic string theory {we can show $I^{--}=I^{++}= \left(-8 \sqrt{2}+O\left(s^2\right)\right)t+O\left(t^2\right)$ and $I^{-+}=I^{+-}=\left(4 \sqrt{2}+O\left(s^3\right)\right)t^2+O\left(t^3\right)$, which are essentially zero in the forward limit.} The bosonic string theory is discussed in the appendix.
\bibitem{foot3}
{The reason for this definition is understood by considering $P(A_1=B_1)$. Obviously this is $P(A_1=0, B_1=0)+P(A_1=1, B_1=1)$.}
\bibitem{foot4}
{Equivalent definition $P\left(A_a=B_b+k\right) \equiv \sum_{j=0}^{d-1} P\left(A_a=j, B_b=j+k \bmod d\right) 
$ can be used with different measurement parameters.}
\bibitem{foot5}
We will use same notation as in \cite{joaophoton}, namely $f_2, f_4, g_4, g_4'$ are normalized w.r.t $g_2$ such that $\bar{f}_i/ ~\bar{g}_i$ are dimensionless.
\bibitem{foot6}
For the S-matrices given in \cite{joaophoton} the value $s=0.01$ means roughly around  $27.7 m_e^2$ if we want to measure this energy in QED. This we obtained as follows. In \cite{joaophoton} the S-matrices are given at $g_2=100$ and we use the formula $g_2=\frac{11\a^2}{45 m_e^4}$, with $\a=1/137$ to figure out the mass of electron to convert $s=0.01$ in units of $m_e^2$.
\end{thebibliography}
\end{document}